\documentclass[12pt,a4paper,tightenlines,nofootinbib]{revtex4}
\usepackage{amssymb}
\usepackage{amsmath}
\usepackage{epsf}
\usepackage{bm}

\newcommand {\si}{\sigma}
\newcommand {\ga}{\gamma}

\newcommand {\de}{\delta}

\newcommand {\be}{\beta}
\newcommand {\pa}{\partial}

\newcommand {\fr}{\frac}
\newcommand {\W}{\hat\Omega}

\newcommand {\z}{\zeta}

\newcommand {\cf}{{\cal F}}

\newcommand {\pe}{\perp}
\newcommand {\lan}{\langle}
\newcommand {\ran}{\rangle}

\newcommand {\bv}{\mathbf{v}}

\newcommand {\baga}{{\gamma}}
\newcommand {\bav}{{v}}

\newcommand {\baw}{{w}}

\newcommand {\baL}{\bar{L}}

\newcommand {\beg}{\begin{equation}}
\newcommand {\en}{\end{equation}}
\newcommand {\bega}{\begin{eqnarray}}
\newcommand {\ena}{\end{eqnarray}}

\begin{document}

\author{Ariel M\'{e}gevand}
\altaffiliation{Member of CONICET, Argentina}
\email[]{megevand@mdp.edu.ar}
\author{Federico Agust\'{\i}n Membiela}
\altaffiliation{Fellow of CONICET, Argentina}
\email[]{membiela@mdp.edu.ar} \affiliation{IFIMAR (CONICET-UNMdP)}
\affiliation{Departamento de F\'{\i}sica, Facultad de Ciencias
Exactas y Naturales,
  UNMdP, De\'{a}n Funes 3350, (7600) Mar del Plata, Argentina}

\title{Stability of cosmological detonation fronts}

\begin{abstract}
The steady state propagation of a phase transition front is
classified, according to hydrodynamics, as a deflagration or a
detonation, depending on its velocity with respect to the fluid.
These propagation modes are further divided into three types, namely,
weak, Jouguet, and strong solutions, according to their disturbance
of the fluid. However, some of these hydrodynamic modes will not be
realized in a phase transition. One particular cause is the presence
of instabilities. In this work we study the linear stability of weak
detonations, which are generally believed to be stable. After
discussing in detail the weak detonation solution, we consider small
perturbations of the interface and the fluid configuration. When the
balance between the driving and friction forces is taken into
account, it turns out that there are actually two different kinds of
weak detonations, which behave very differently as functions of the
parameters. We show that  the branch of stronger weak detonations are
unstable, except very close to the Jouguet point, where our approach
breaks down.
\end{abstract}

\maketitle

\section{Introduction}

In a cosmological first-order phase transition, bubbles of the stable
phase nucleate and expand in the supercooled metastable phase. The
motion and collisions of bubble walls are associated with the
formation of cosmological remnants, such as gravitational waves
\cite{gw}, magnetic fields \cite{gr01}, topological defects
\cite{vs94}, a baryon asymmetry \cite{ckn93}, or baryon
inhomogeneities \cite{w84,h95,ma05}. The propagation of bubble walls
is driven essentially by the pressure difference between the two
phases, but is significantly affected by hydrodynamics \cite{hidro}
(for recent studies, see \cite{ms09,ekns10,lm11,kn11,ms12}). Indeed,
besides stirring the fluid, these phase transition fronts reheat the
plasma, due to the release of latent heat. The back-reaction of such
disturbances hinders the wall motion. In addition, the microscopic
interaction of particles of the plasma with the wall causes, at the
macroscopic level, a friction force \cite{micro,bm09,hs13,ariel13}.
In general, after a time which is much shorter than the total
duration of the phase transition, the wall reaches a steady state
with constant velocity (however, the wall may also run away
\cite{bm09}).

As a consequence of nonlinear hydrodynamics, different kinds of
stationary solutions exist. Thus, the phase transition front may
propagate, in principle, either as a weak, Jouguet, or strong
deflagration, as well as a weak, Jouguet, or strong detonation. The
latter, however, is not possible, since its fluid profile cannot
fulfil the boundary conditions. These various propagation modes
coexist in some ranges of parameters, and it is not easy to
determine, in general, which of them will be actually reached in the
phase transition. It will certainly depend on the initial and
boundary conditions during the early transitory stage.

In the literature, the wall velocity is often given convenient values
or left as a free parameter. For instance, slow-moving weak
deflagrations are often assumed for baryogenesis, while fast-moving
detonations are assumed for gravity-wave generation. In the latter
case, a Jouguet detonation is often assumed. This solution
corresponds to the lower bound for the detonation velocity. The
choice of this particular solution avoids calculating the wall
velocity, which would involve considering the balance of the driving
and friction forces. This assumption is motivated by the case of
chemical burning, where weak detonations are forbidden  \cite{s82}.
However, a phase transition is qualitatively different from chemical
burning, and weak detonations are possible in addition to Jouguet
detonations \cite{l94}.

Even when a complete calculation of the wall velocity is performed,
there still remains the problem of the existence of multiple
solutions (for a recent discussion, see \cite{ms12}). Besides the
coexistence of the aforementioned kinds of solutions,  a calculation
of the wall velocity may give double-valued solutions of a given
kind. In particular, for the weak detonation case, there are in
general two branches. One of these branches consists of higher
velocity solutions, which are hydrodynamically weaker (they cause
smaller disturbances of the fluid). These solutions approach the
speed of light for low enough friction or high enough supercooling.
The other branch of weak detonations consists of stronger solutions
with lower velocities, and ends at the Jouguet point. These
detonations behave unphysically as a function of the parameters.
Therefore, weak detonations might be divided into ``weaker'' weak
detonations and ``stronger'' weak detonations.

An important issue is, thus, to determine the actual propagation mode
for a given set of parameters. A stability analysis provides a useful
tool to determine the final state of the wall motion, since not all
of the possible hydrodynamic solutions turn out to be stable. For
instance, it is known that strong deflagrations are generally
unstable, and that weak deflagrations are unstable in certain ranges
of parameters \cite{hkllm,stabdefla}. The case of detonations is less
clear.

In the cosmological context, the stability of weak detonations was
first discussed by Huet et al. \cite{hkllm}. The standard approach is
to consider small perturbations of the fluid variables on each side
of the wall, together with small deformations of the latter, which is
assumed to be infinitely thin \cite{landau,link}. The fluid
perturbations on the two phases are linked by junction conditions at
the interface. In the case of detonations, the fluid enters the wall
supersonically (in the reference frame of the latter). As a
consequence, fluid perturbations cannot evolve in front of the wall.
This fact lead Huet et al. to conclude that perturbations cannot grow
at all, and weak detonations are always stable. However, numerical
calculations of the time evolution of the wall-fluid system seem to
indicate that configurations belonging to the ``stronger'' branch of
weak detonations are unstable \cite{ikkl94}.

In fact, as noted by Abney \cite{abney}, the conclusion of Ref.
\cite{hkllm} is incorrect, as fluid perturbations may vanish in front
of the wall and not behind it (and fulfil the junction conditions).
Such perturbations, which were not considered in \cite{hkllm}, may be
unstable. Unfortunately, Abney (and later Rezzola \cite{r96})
considered only Jouguet and strong detonations. As already mentioned,
the latter are not possible in a phase transition. Moreover, even in
the case of Jouguet detonations, the treatment of Refs.
\cite{abney,r96} is not valid. In the first place, an equation for
the interface (depending on the driving and friction forces) was not
considered in these works, even though its importance had been
already pointed out be Huet et al. \cite{hkllm}. Most important, the
treatment considers perturbations around a solution of constant fluid
velocity $v$ and temperature $T$, whereas the fluid profile for a
Jouguet detonation is that of a rarefaction wave behind the wall,
with $v$ and $T$ varying in space and time.

In this paper we shall consider the hydrodynamic stability of
detonation fronts in a cosmological phase transition. We shall
consider only weak detonations. Unfortunately, for the case of
Jouguet detonations, the fact that the fluid profiles are not
constant behind the wall makes the treatment too involved. For the
same reason, our approach will break down for weak detonations which
are very close to the Jouguet point. Except in this limit, we will
show that the lower velocity branch of weak detonations is unstable
under linear perturbations at all wavelengths. Our approach is
essentially the same we used in Ref. \cite{stabdefla} for the case of
weak deflagrations, and we shall refer to that work for some details.
Due to the vanishing of perturbations in the metastable phase, the
treatment of detonations turns out to be much simpler than that of
deflagrations. This will allow us to obtain analytical and
model-independent results.

The plan of the paper is the following. Before performing the
stability analysis, we devote the next section to a detailed
discussion on the structure of weak detonations. We also consider the
coexistence with runaway solutions. In Sec. \ref{staban} we solve the
equations for the linear perturbations of the wall-fluid system, and
in Sec. \ref{stab} we discuss in detail the velocity intervals
corresponding to unstable solutions and to the validity of our
approach. We also discuss some implications of our results for a
cosmological phase transition. Our conclusions are summarized in Sec.
\ref{conclu}.

\section{Detonation fronts in a phase transition} \label{stationary}

For a given theory, a phase transition may occur if the free energy
depends on a scalar field $\phi$ which acts as an order parameter
\cite{quiros}. Thus, if the free energy density $\mathcal{F}(\phi,T)$
has a minimum $\phi_+$ at high temperature, and a different minimum
$\phi_-$ at low temperature, we have two different phases. Each phase
will be described by a free energy density $ \mathcal{F}_{\pm}(T)=
\mathcal{F}(\phi_\pm,T)$. Indeed, all the thermodynamical quantities
in each phase can be derived from the functions
$\mathcal{F}_{\pm}(T)$. Such a relation between thermodynamical
quantities is known as an equation of state (EOS). For instance, the
pressure is given by $p=-\mathcal{F}(T)$, the entropy density by
$s=dp/dT$, the enthalpy density by $w=Ts$, and the energy density by
$e=Ts-p$.

If the phase transition is first-order, there is a range of
temperatures at which the two minima $\phi_\pm$ coexist, separated by
a barrier. The critical temperature $T_c$ is that at which $\phi_+$
and $\phi_-$ have the same free energy, i.e., $T_c$ is defined by
$\mathcal{F}_{+}(T_{c})=\mathcal{F}_{-}(T_{c})$. Some quantities
(e.g., the energy and the entropy) are discontinuous at $T=T_c$. The
latent heat is defined as the energy density discontinuity at
$T=T_c$, and is given by $L
=T_c[\mathcal{F}_-'(T_c)-\mathcal{F}_+'(T_c)]$.

In the early universe, the system is initially in the
high-temperature phase. As the temperature descends below $T_c$, the
$+$ phase becomes metastable, but the system remains in this
supercooled phase due to the free-energy barrier between minima (see,
e.g., \cite{gw81,linde,ah92}). Finally, at some temperature
$T_N<T_c$, bubbles of the stable low-temperature phase begin to
nucleate and grow, until they fill all space (for reviews on phase
transition dynamics, see, e.g., \cite{mege}). The expectation value
of the field takes the value $\phi_-$ inside the bubble and the value
$\phi_+$ outside it. Thus, the bubble can be seen as a
classical-field configuration. The walls of these bubbles interact
with the particles of the plasma. At the same time, the bubble walls
are phase transition fronts, at which latent heat is released.

Macroscopically, we can describe the field-fluid system by
considering the conservation of the stress tensor, together with a
finite-temperature equation for the field. These equations can be
written in the form
  \bega
      \pa_\mu\left(-T\fr{\pa\cf}{\pa T}u^\mu u^\nu+g^{\mu\nu}
      \cf\right)+\pa_\mu\pa^\mu\phi\pa^\nu\phi=0\label{fluideq},\\
      \pa_\mu\pa^\mu\phi+\fr{\pa\cf}{\pa\phi}+\fr{\tilde\eta\,
      u^\mu\pa_\mu\phi}{\sqrt{1+(\tilde{\lambda}\, u^\mu\pa_\mu\phi)^2}}=0\label{fieldeq},
  \ena
with $u^\mu=(\ga,\ga\bv)$ the four velocity of the fluid and
$g^{\mu\nu}$ the Minkowsky metric tensor. Notice that the terms in
parenthesis in Eq. (\ref{fluideq}) correspond to the familiar stress
tensor of the relativistic fluid, $ wu^\mu u^\nu-p g^{\mu\nu}$. The
last term in this equation gives the transfer of energy between the
plasma and the field. The field is governed by Eq. (\ref{fieldeq}),
where we have included, as is customary, a phenomenological damping
term proportional to $u^\mu\pa _\mu \phi$. This term will give the
friction force between the wall and the plasma. The particular form
used in Eq. (\ref{fieldeq}) was proposed in Ref. \cite{ariel13} in
order to account for the saturation of the friction force at
ultra-relativistic velocities \cite{bm09}. The coefficients
$\tilde\eta$ and $\tilde\lambda$ can be obtained from microphysics
calculations (in the general case, $\tilde\eta$ and $\tilde\lambda$
may depend on the field $\phi$). The coefficient $\tilde\eta$ will
dominate for non-relativistic fluid velocities, whereas the
coefficient $\tilde \lambda$ will dominate the ultra-relativistic
behavior. Similar damping terms (with a single parameter) have been
considered in Refs. \cite{ekns10,hs13}.

For the distance scales associated to the wall motion and
hydrodynamic profiles, the bubble wall can be regarded as an
infinitely thin interface. Due to the friction with the plasma, this
interface will in general reach a terminal velocity. However, due to
the saturation of the friction force at ultra-relativistic
velocities, a state of continuous acceleration \cite{bm09} also
exists. Besides, as we have already mentioned, stationary solutions
may be unstable. We shall now consider the stationary motion of a
planar interface, focusing on the detonation case, and in the next
section we shall study perturbations of this configuration.

\subsection{Hydrodynamic solutions}

\begin{figure}[bth]
\centering
\epsfysize=2cm \leavevmode \epsfbox{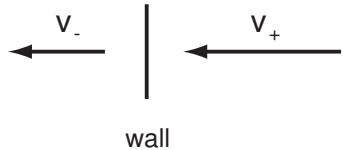}
\caption{Sketch of a detonation in the wall frame.}
\label{figwall}
\end{figure}

Let us consider a planar wall moving towards the positive $z$ axis.
Due to the planar symmetry, the fluid velocity is perpendicular to
the wall (see Fig. \ref{figwall}). The fluid velocity and the
temperature are different on each side of the wall. The relation
between these values can be obtained by integrating Eq.
(\ref{fluideq}) across the interface. In the reference frame of the
wall, we obtain
 \bega
    w_-v_-\ga^2_-&=&w_+v_+\ga^2_+,\label{EM3.1}\\
    w_-v_-^2\ga^2_- +p_-&=&w_+v^2_+\ga^2_+ +p_+ , \label{EM3.2}
 \ena
where $v$ is the $z$ component of the fluid velocity, and $+$ and $-$
signs refer to variables just in front and just behind the wall,
respectively. Notice that in this frame we have $v_{\pm}<0$.
If the equation of state is known, Eqs. (\ref{EM3.1}-\ref{EM3.2}) can
be used to obtain the velocity and temperature of the outgoing flow
as functions of the incoming flow variables. Nevertheless, some
general features of these hydrodynamic relations do not depend on the
EOS.

In the first place, it is easy to see that we have two branches of
hydrodynamic solutions. Indeed, from Eqs. (\ref{EM3.1}-\ref{EM3.2})
we may write
 \beg
\fr{v_+-v_-}{v_+}=\fr{p_--p_+}{w_+\ga_+^2 v_+^2}.
 \en
Therefore, we see that, for given values of $v_+$ and $T_+$, we will
have in general {two solutions} for the outgoing flow. One of them
with\footnote{Notice that $(v_+-v_-)/v_+=(|v_+|-|v_-|)/|v_+|$.}
$|v_-|<|v_+|$ and $p_+<p_-$, called \emph{detonation}, and the other
with $|v_-|>|v_+|$ and $p_+>p_-$, called \emph{deflagration} (the
sketch of Fig. \ref{figwall} corresponds to a detonation
configuration).

In the second place, notice that \emph{variations} of thermodynamical
quantities are generally related by the speed of sound, which is
given by $c_s^2\equiv dp/de$. We have, in particular,
 \beg
 dp=dw/(1+c_s^{-2}). \label{pwcs}
 \en
Therefore, differentiating Eqs. (\ref{EM3.1}-\ref{EM3.2}) we may
gather all the information on the EOS in this single parameter. Let
us regard the temperature $T_+$ in front of the wall as a boundary
condition (this is the case if the incoming flow is supersonic), and
consider the dependence of $v_-$ and $T_-$ on $v_+$. We have
 \bega \label{dif1}
 \left(v_-^2\ga_-^2+\fr{c_{s-}^2}{1+c_{s-}^2}\right)dw_- +2w_+v_+\ga_+^2\ga_-^2dv_-=2w_+v_+\ga_+^4dv_+,\\
 v_-\ga_-^2 dw_-
 +w_+\ga_+^2\ga_-^2\fr{v_+}{v_-}(1+v_-^2)dv_-=w_+\ga_+^4(1+v_+^2)dv_+,
 \label{dif2}
 \ena
where we have used the relations $d(v^2 \ga^2)=2v\ga^4dv$,
$d(v\ga^2)=\ga^4(1+v^2)dv$. Combining Eqs. (\ref{dif1}-\ref{dif2}),
we obtain
 \bega
\left.\fr{\pa v_-}{\pa v_+}\right|_{T_+}&=&
      \fr{\ga_+^2}{\ga_-^2}+
      \fr{\ga_+^2(v_-^2+c_{s-}^2)(1-v_+v_-)}{{v_-^2-c_{s-}^2}}\fr{v_+-v_-}{v_+},
      \label{dvmedvma} \\
\fr{1}{w_-}\left.\fr{\pa w_-}{\pa v_+}\right|_{T_+}&=& -
      \fr{2v_-\ga_-^2\ga_+^2(1+c_{s-}^2)(1-v_+v_-)}{v_-^2-c_{s-}^2}\fr{v_+-v_-}{v_+}.
      \label{dwmedvma}
 \ena
Notice that these equations do not depend on the sign of $v_\pm$, and are
valid for a wall propagating towards the negative $z$ axis as well.

\begin{figure}[bth]
\centering
\epsfysize=5cm \leavevmode \epsfbox{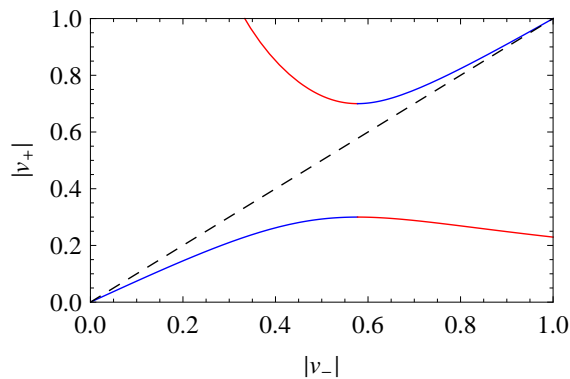}
\caption{The fluid velocities in the rest frame of the wall for fixed $T_+$,
obtained from Eq. (\ref{dvmedvma}) for $c_{s_-}=1/\sqrt{3}$,
with  conditions $v_+(c_{s-})=0.3$ (lower curve) and $v_+(c_{s-})=0.7$ (upper curve).
Blue lines indicate weak solutions and red lines indicate strong solutions. The dashed line
indicates the value of $v_+=v_-$.}
\label{figvmavme}
\end{figure}

It is evident that the speed of sound in the $-$ phase plays a
relevant role in Eqs. (\ref{dvmedvma}-\ref{dwmedvma}). For instance,
the inverse of the derivative in Eq. (\ref{dvmedvma}) vanishes at
this point, which means that he curve of $v_+$ as a function of $v_-$
has an extremum. Thus, the case $|v_-|= c_{s-}$ separates two
different behaviors, namely, $v_+$ growing with $v_-$, or $v_+$
decreasing with $v_-$. The former are called weak solutions and the
latter are called strong solutions. The point $|v_-|= c_{s-}$ is
called the Jouguet point. The extremum of $v_+$ vs. $v_-$ at the
Jouguet point will be either a maximum or a minimum, depending on the
sign of $(v_+-v_-)/v_+$. Thus, for detonations we have a minimum,
whereas for deflagrations we have a maximum (see Fig.
\ref{figvmavme}).

Physically, we know that there must be solutions with low values of
$v_-$ and $v_+$, corresponding to a slow wall. In the limit of a
vanishingly small wall velocity, we must have $v_-=v_+=0$, while if
the wall velocity is small but nonvanishing, both $v_-$ and $v_+$
will be nonvanishing (and will have the same sign). Hence, these
solutions correspond to the \emph{weak} part of a curve (i.e., $\pa
v_+/\pa v_->0$). If we increase further the velocity, we will reach
the Jouguet point $|v_-|=c_{s-}$, where, according to the above,
$|v_+|$ has an extremum. In this case, the extremum must be a
maximum, and $|v_+|$ will decrease for $|v_-|>c_{s-}$ (strong
solutions). Thus, we are in a deflagration curve.

On the other hand, there must also be solutions with $v_-\approx
v_+\approx 1$, corresponding to a very fast moving wall. As the wall
velocity decreases from the limit $v_w=1$, both $|v_+|$ and $|v_-|$
will decrease. Therefore, we are again in the weak part of a curve.
This curve must have a minimum at the Jouguet point $|v_-|=c_{s-}$
and, thus, corresponds to detonation solutions.

Thus, we see that, for detonations, the incoming flow is supersonic,
whereas for deflagrations the incoming flow is subsonic. Notice also
that weak solutions correspond to smaller values of $v_+-v_-$ than
strong solutions and, thus, to weaker disturbances of the fluid. For
weak deflagrations, the wall velocity is subsonic with respect to the
fluid on both sides of it, whereas for weak detonations the wall
velocity is supersonic with respect to the fluid on both sides.

We may obtain the form of the detonation and deflagration curves if
we integrate Eqs. (\ref{dvmedvma}-\ref{dwmedvma}), although some
information will be lost with respect to Eqs.
(\ref{EM3.1}-\ref{EM3.2}) (namely, there will be undetermined
integration constants). In general, $c_{s-}$ is a function of $w_-$,
and the two derivatives in (\ref{dvmedvma}-\ref{dwmedvma}) cannot be
integrated independently. Nevertheless, if we neglect the variation
of the speed of sound, Eq. (\ref{dvmedvma}) can be integrated alone.
In Fig. \ref{figvmavme} we show the result\footnote{We actually
integrated the inverse of Eq. (\ref{dvmedvma}), i.e., $\pa v_+/\pa
v_-$.} for the case $c_{s-}=1/\sqrt{3}$ (corresponding to an
ultra-relativistic gas). We have arbitrarily chosen two conditions to
determine the integration constant, corresponding to detonations and
deflagrations (for a specific EOS, the two values of $v_+$ for a
given $v_-$ will be determined as functions of $T_+$). The result is
similar for any constant value of $c_{s-}$. In the general case the
curves are also similar, but $c_{s-}$ may be different in each curve,
as it depends on the temperature.

\subsection{Fluid profiles} \label{hydrostat}

Away from the wall, the field is a constant and Eq. (\ref{fluideq})
gives the conservation of energy and momentum for the fluid, i.e.,
$\pa_\mu (wu^\mu u^\nu-p g^{\mu\nu})=0$. Using Eq. (\ref{pwcs}), we
may obtain again  equations which depend only on the parameter $c_s$
\cite{landau}. Besides, the absence of a distance scale in these
equations justifies to assume the \emph{similarity condition},
namely, that the solution will depend only on the variable $\xi=z/t$.
In the planar case, one obtains very simple equations for $v(\xi)$
and $w(\xi)$ (see e.g. \cite{lm11}). The solutions are\footnote{In
fact, there is also a solution $v=(\xi+c_{s})(1+\xi c_{s})$, but this
solution will not fulfil the matching and boundary conditions
\cite{lm11}.}
 \beg
 v(\xi)= \mathrm{constant}
 \en
and the rarefaction wave
 \beg
 v_{\mathrm{rar}}(\xi)=\fr{\xi-c_{s}}{1-\xi c_{s}}. \label{rar}
 \en
Notice that there is no integration constant in Eq. (\ref{rar}). The
corresponding solutions for the thermodynamical variables are given
by
 \beg
 w(\xi)= \mathrm{constant}
 \en
and by
 \beg
\fr{c_s}{1+c_s^2}\fr{dw}{w}=\gamma^2dv.
 \en
The latter is trivially integrated in the case of constant $c_s$.

The boundary conditions are, in the reference frame of the bubble
center, that the fluid velocity vanishes far behind the wall (i.e.,
at the center of the bubble) and far in front of the wall (where the
fluid is still unperturbed). Therefore, in the wall frame, the fluid
velocity far in front and far behind the wall must be given by
$v=-v_w$. The boundary condition for the temperature is that its
value far in front of the wall is given by the nucleation temperature
$T_N$.

\begin{figure}[hbt]
\centering
\epsfysize=5cm \leavevmode \epsfbox{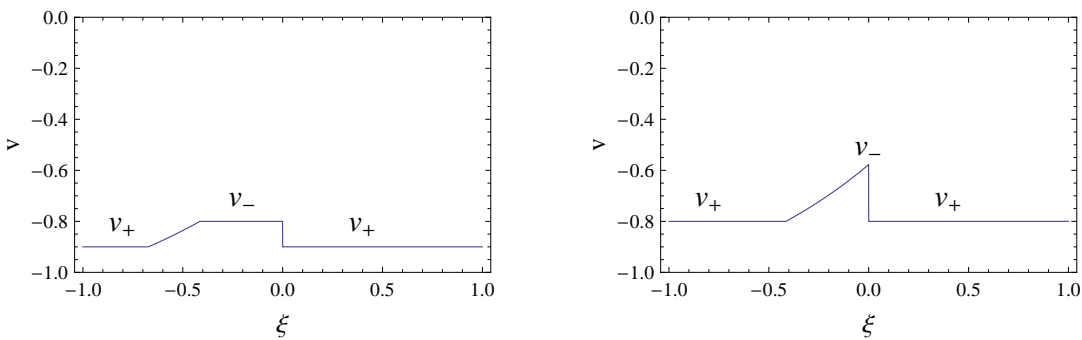}
\caption{The fluid velocity profile for $c_{s-}=1/\sqrt{3}$,
in the reference frame of the bubble wall. Left panel: a
weak detonation ($v_+=-0.9$, $v_-=-0.8$). Right panel:
a Jouguet detonation ($v_+=-0.8$, $v_-=-c_{s-}$).}
\label{figdeto}
\end{figure}

It is not difficult to construct the fluid  velocity and temperature
profiles from the similarity solutions, using the boundary conditions
and the matching conditions at the wall. Let us consider the
detonation case. As we have seen, the incoming flow is supersonic
and, hence, has not received any information from the wall. As a
consequence, the boundary condition gives $v_+=-v_w$. As we have
seen, for a detonation we have  $|v _-|<|v_+|$ (hence, $v_->v_+$),
which means that the fluid  in the $-$ phase is dragged by the wall.
Behind the wall, the velocity must descend
continuously\footnote{Entropy considerations show that we cannot have
a discontinuity with $v$ increasing in the direction of the front
propagation (for details, see e.g. \cite{lm11}).} from its value at
the wall, $v_-$, to the value at the bubble center, $-v_w$. To
achieve this, we must use the rarefaction solution (\ref{rar}) as
well as constant solutions $v=v_+$ and $v=v_-$, as shown in the left
panel of Fig. \ref{figdeto}\footnote{In the figure we have chosen
arbitrarily the values of $v_+$ and $v_-$. For a given EOS, these
values will be given by the matching conditions.}. In the wall frame,
the wall is at $\xi_w=0$. The rarefaction solution matches the value
$v_-$ at
 \beg
 \xi_0=(c_{s-}+v_-)/(1+c_{s-}v_-) \label{xi0}
 \en
and the value $v_+$ at
 \beg
 \xi_c=(-v_w+c_{s-})/(1-v_wc_{s-})
 \en
In the frame of the bubble center, the fluid velocity vanishes at
this latter point, which moves with velocity $c_{s-}$. The
temperature profile is qualitatively similar to the velocity profile.
In front of the wall we have a constant temperature $T_+=T_N$. Behind
the wall the fluid is reheated to a temperature $T_-$, which then
descends along the rarefaction wave to a final value $T_f\simeq T_N$.

Notice that this profile is only possible for weak or Jouguet
detonations (i.e., for $v_-\leq -c_{s-}$). In fact, at the Jouguet
point we have $\xi_0=\xi_w=0$, and the region of constant velocity
$v_-$ disappears (right panel of Fig. \ref{figdeto}). A strong
detonation would require $\xi_0>\xi_w$, and the fluid profile cannot
be constructed. Therefore, strong detonations are not possible.

Strong deflagration profiles can be constructed, but are unstable
(see \cite{stabdefla} for details). For a weak deflagration, both
$v_+$ and $v_-$ are subsonic, and the rarefaction solution
$v_{\mathrm{rar}}(\xi)$ cannot be accommodated in the velocity
profile. Therefore, the fluid velocity takes the constant value
$v=v_-$ everywhere behind the wall. According to the boundary
condition at the bubble center, we thus have  $v_-=-v_w$. For a
deflagration we have $|v_+|<|v_-|$, which means that the fluid is
pushed forward in front of the wall. Hence, we have a constant
solution $v=v_+>v_-$ up to a certain point $\xi_{\mathrm{sh}}$, where
the velocity must descend abruptly to take the boundary value
$v=-v_w=v_-$. Such a discontinuity without a change of phase is
called a shock front. At the shock discontinuity, Eqs.
(\ref{EM3.1}-\ref{EM3.2}) still apply, but now the enthalpy and
pressure are related by the same EOS on both sides of the interface.
These equations give the temperature $T_+$ as a function of the
boundary value $T_N$, as well as the velocity of the shock front. The
fluid is reheated in front of the wall, i.e., $T_+>T_-,T_N$.

Notice that the velocity of a weak deflagration is in the range
$0<v_w<c_{s-}$. On the other hand, the velocity of a weak detonation
is in the range $v_J<v_w<1$, where $v_J$ is the Jouguet detonation
velocity, corresponding to the value of $|v_+|$ at $|v_-|=c_{s-}$ for
the detonation curve. The latter is supersonic (see Fig.
\ref{figvmavme}). The velocity gap between $c_{s-}$ and $v_J$ is
filled by a family of supersonic Jouguet deflagrations. These
solutions fulfil the Jouguet condition $v_-=-c_{s-}$, as well as the
deflagration condition $|v_+|<|v_-|$, i.e., the fluid comes in
subsonically, and goes out at the speed of sound. Nevertheless, the
fluid on both sides moves forward with the wall, and the wall
velocity with respect to the bubble center is not $c_{s-}$ but
higher. The velocity profile behind the wall is as in the case of the
Jouguet detonation (right panel of Fig. \ref{figdeto}). On the other
hand, in front of the wall we have $v_+>v_-$, and the profile shows a
shock front discontinuity (for details, see, e.g., \cite{lm11}).

It is important to stress that, for a given value of $T_N$, the
Jouguet deflagrations are a whole set of solutions with velocities
ranging from $v_w=c_{s-}$ to $v_w=v_J$ (both values depend on $T_-$,
which, in turn, depends on $T_N$). In contrast, the Jouguet
detonation is a single solution, corresponding to the lowest possible
value of the detonation velocity, $v_w=v_J$.

From the above, it is apparent that hydrodynamics alone does not
determine the value of the wall velocity. An additional equation is
needed, corresponding to the balance between the driving and friction
forces.

\subsection{Equation for the interface}

In order to obtain a macroscopic equation for the bubble wall, we
need to consider the microphysics inside this  thin interface, which
is described by Eq. (\ref{fieldeq}). Consider a reference frame at
the center of the wall (thus, the field profile only varies in a
small region around $z=z_w=0$). In the steady state, this frame moves
at constant velocity, and only $z$ derivatives appear in Eq.
(\ref{fieldeq}). We multiply by $\phi'\equiv d\phi/dz$ and then we
integrate across the wall (notice that $\phi'$ vanishes outside the
wall). Using the relation $(\pa \cf / \pa \phi) d\phi=d\cf -
(\pa\cf/\pa T)dT$, we obtain \cite{stabdefla,ariel13}
\begin{equation}
\sigma \ddot{z}_w=F_{\mathrm{dr}} + F_{\mathrm{fr}}, \label{fuerzas}
\end{equation}
where $\sigma$ is the surface tension,
 \beg
 \sigma\equiv\int\phi'^2dz,
 \en
$F_{\mathrm{dr}}$ is the force (per unit area) driving the
propagation of the phase transition front,
 \beg
   F_{\mathrm{dr}} =  p_-(T_-)-p_+(T_+)+\int_{T_-}^{T_+}
   \frac{\partial\mathcal{F}}{\pa T}{dT},  \label{fdrexac}
 \en
and $F_{\mathrm{fr}}$ is the friction force,
\begin{equation}
F_{\mathrm{fr}}=\int \frac{\gamma v\, \tilde\eta(\phi')^2}{ \sqrt{1+
(\gamma v)^2 \,\tilde\lambda^2 (\phi')^2}}\,dz , \label{feno}
\end{equation}
which  depends explicitly on the velocity of the wall with respect to
the fluid, $-v$ (notice that $F_{\mathrm{fr}}$ is negative since we
have $v<0$).

In order to obtain  macroscopic expressions which do not depend on
the wall shape, we will approximate the integrals in Eqs.
(\ref{fdrexac}) and (\ref{feno}). For the integral (\ref{feno}),  we
notice that the function $\phi'(z)^2$ peaks inside the wall.
Therefore, its presence in the numerator will select values of the
integrand around $z\approx 0$. Hence, inside the square root in the
denominator, we will just approximate $\phi'^2$ by its value at the
center of the wall, $\phi'^2_0$, whereas in the numerator, we will
approximate it by a delta function $\phi'^2_0 l_w\delta(z)$, where
$l_w$ is the wall width. Furthermore, we may approximate the value of
$\phi'^2_0$ by $(\Delta\phi/l_w)^2$, where $\Delta \phi$ is the field
variation $\phi_+-\phi_-$.  The details of the approximations are not
relevant, since all these parameters will be absorbed in the free
parameters $\tilde \eta$ and $\tilde \lambda$. Indeed, we define
$\eta= \tilde\eta (\Delta\phi)^2/l_w$,  $\lambda= \tilde\lambda
\Delta\phi/l_w$. With these approximations, the integrand now depends
only on $v$, which we assume has a smooth variation between $v_-$ and
$v_+$. We thus have an integral of the form $\int
f(z)\delta(z)dz=f(0)$. Since we do not know the exact value of
$f(0)$, we shall approximate it by the average of its values on each
side of the wall,  $\langle f \rangle\equiv(f_++f_-)/2$ . We thus
obtain \cite{ariel13}
\begin{equation}
F_{\mathrm{fr}}=\left\langle\frac{\eta \,\gamma v }{ \sqrt{1+
\lambda^2(\gamma v)^2}}\right\rangle
. \label{fric}
\end{equation}%
For a given model, the values of $\eta$ and $\lambda$ can be obtained
by comparing the non-relativistic and ultra-relativistic limits of
Eq. (\ref{fric}) with the corresponding results from microphysics
calculations \cite{ariel13}. Indeed, notice that, for small $v_w$, we
have $F_{\mathrm{fr}}=-\eta v_w$, while for $v_w\to 1$ we have
$F_{\mathrm{fr}}=-(\eta/\lambda) v_w$.

For the integral in Eq. (\ref{fdrexac}), we may just use a linear
approximation for the integrand inside the wall. It is useful to use,
first, the identity $(\pa \mathcal{F}/\pa T)dT=(\pa \mathcal{F}/\pa
T^2)dT^2$ to obtain an expression in terms of $T^2$. This is
convenient since $\mathcal{F}$ is often quadratic. Thus, we obtain,
 \beg
F_{\mathrm{dr}}=p_-(T_-)-p_+(T_+)
+\left\langle\fr{dp}{dT^2}\right\rangle\left(T_+^2-T_-^2\right).
\label{fdr2}
 \en
The second term in Eqs. (\ref{fdrexac}) and (\ref{fdr2}) is due to
hydrodynamic effects. This term is very important, since the driving
force cannot be exclusively determined by the pressures outside the
wall. For instance, for a deflagration, as we have seen, we have
$p_-(T_-)-p_+(T_+)<0$.

Since the friction force increases with the velocity, the wall may
reach a terminal velocity, given by the equation
$F_{\mathrm{fr}}=-F_{\mathrm{dr}}$. Such a stationary state will be
reached in a time which is very short in comparison to the duration
of a phase transition. Indeed, according to Eq. (\ref{fuerzas}), the
time scale associated to the acceleration of the wall is given by the
quantity
\begin{equation}
d=\frac{\sigma}{F_{\mathrm{dr}}} \label{d}.
\end{equation}
Since $\sigma$ and $F_{\mathrm{dr}}$ are determined by the scale of
the phase transition, this quantity is roughly given by $d\sim 1/T$,
while the duration of the phase transition is roughly given by the
cosmological time scale $t\sim M_P/T^2$, where $M_P$ is the Plank
mass. Notice that the ratio in Eq. (\ref{d}) gives also a length
scale associated to the balance between the driving force and the
surface tension, in the case of a deformed wall (see the next
section).

If a stationary state is reached, the terminal velocity is given by
the equation
 \beg
   \eta\left\langle{\gamma_{\lambda} v }\right\rangle =- F_{\mathrm{dr}},
   \label{statv}
 \en
where we have rewritten the friction force (\ref{fric}) in terms of
the quantity
 \beg
 \gamma_\lambda=\fr{1}{\sqrt{1-(1-\lambda^2)v^2}}.
 \en
This equation can be solved using Eqs. (\ref{EM3.1}-\ref{EM3.2}) and
appropriate boundary conditions.

\subsection{Runaway walls}

In the particular case $\lambda=0$, we have $\gamma_\lambda =\gamma$,
and the friction force grows as $\gamma v$. In such a case, the wall
will always reach a terminal velocity $v_w<1$. On the other hand, for
nonvanishing $\lambda$, the friction saturates at the value
$F_{\mathrm{fr}}=-\eta/\lambda$ in the ultra-relativistic limit. If
the driving force (which depends on the amount of supercooling and on
hydrodynamics) is larger than this value, a terminal velocity will
not be reached, i.e.,  the wall will run away. The condition for
non-existence of stationary solutions is thus
 \beg
F_{\mathrm{dr}}>\eta/\lambda , \label{nostat}
 \en
with $F_{\mathrm{dr}}$ given by Eq. (\ref{fdr2}). Notice that the
condition  (\ref{nostat}) can be reached only in the
ultra-relativistic limit. Therefore, as a function of the parameters,
Eq. (\ref{nostat}) is a condition for the nonexistence of
\emph{detonations}. Indeed, for a given set of parameters, the
detonation solution may fulfill Eq. (\ref{nostat}) (which means that
its velocity would exceeded the speed of light), but it may happen
that we  still have a deflagration solution, for which the driving
force is much smaller (due to different hydrodynamics). Hence,
deflagrations may coexist with runaway solutions, even when
detonations do not.

Detonations may also coexist with runaway solutions. Indeed, a
runaway solution may exist even if Eq. (\ref{nostat}) is not
fulfilled. This is possible because, for extremely high values of
$\gamma v$, the hydrodynamics is different and  $F_{\mathrm{dr}}$ is
not given by Eq. (\ref{fdr2}) anymore \cite{bm09,ariel13}. In this
case, the particles distribution functions are essentially unaffected
as they pass through the wall. The \emph{total} force in this limit
is given by the difference
$F_{\mathrm{tot}}=\mathcal{F}_+(T_+)-\tilde{\mathcal{F}}_-(T_+)$,
where $\tilde{\mathcal{F}}_-(T_+)$ is the mean field effective
potential, obtained by keeping only the quadratic terms in a Taylor
expansion about the $+$ phase \cite{bm09,ekns10}. In our
phenomenological model, this total force is given by
$F_{\mathrm{tot}}=p_-(T_+)-p_+(T_+)-\eta/\lambda$ (see \cite{ariel13}
for details). Therefore, the condition for the existence of a runaway
solution is
 \beg \label{runaway}
 p_-(T_+)-p_+(T_+)>\eta/\lambda.
 \en

For given values of the friction parameters, both conditions
(\ref{nostat}) and (\ref{runaway}) will be fulfilled for high enough
supercooling (i.e., small enough values of $T_+/T_c$). Nevertheless,
it is important to notice that, when the condition (\ref{runaway}) is
already fulfilled, the condition (\ref{nostat}) may not be fulfilled
yet. Indeed, for a stationary solution the driving force is smaller
than $p_-(T_+)-p_+(T_+)$ (due to hydrodynamic effects). Hence, even
if the runaway solution exists, the wall may still not run away,
since a detonation solution may exist as well. In the case of
coexistence of a stationary solution and a runaway solution, one
expects that the former will be the one to be realized in the phase
transition, unless it is unstable.

\subsection{A specific example: the bag EOS}

The simplest phenomenological equation of state for a phase
transition is the well known bag EOS, which consists of radiation and
vacuum energy. The pressure in each phase can be written in the form
 \beg
 p_+(T)=\fr{a}{3}T^4-\fr{L}{4},\;\;   p_-(T) =\left(\fr{a}{3}-\fr{L}{4T_c^4}\right)T^4.
 \en
The thermodynamic quantities can be obtained from $s=dp/dT,w=Ts,
e=Ts-p$. This model has three free parameters, namely, the critical
temperature $T_c$, the latent heat $L$, and the coefficient $a$. The
latter is related to the number of degrees of freedom of radiation in
the $+$ phase. The speed of sound is the same in both phases,
$c_{s}=1/\sqrt{3}$.

In this model, the driving force (\ref{fdr2}) takes the form
 \beg
 F_{\mathrm{dr}}=\fr{L}{4}\left(1-\fr{T_-^2T_+^2}{T_c^4}\right).
 \label{fdrbag}
 \en
The matching conditions (\ref{EM3.1}-\ref{EM3.2}) give the relations
 \beg \label{tme2}
 \fr{T_-^2}{T_+^2}=\sqrt{\fr{v_+ \gamma_+^2}{v_- \gamma_-^2 (1 -
 \baL)}},
 \en
\begin{equation}
v_{-} =\fr{1}{6v_+}\left[ 1-3\alpha + 3 \left(1+\alpha \right)v_{+}^2
\pm \sqrt{\left[1-3\alpha + 3\left(1+\alpha \right)  v_{+}^2 \right] ^{2}-12v_+^2}\right] ,
\label{steinhardt}
\end{equation}
where
 \beg
 \baL\equiv \fr{L}{4aT_c^4/3}=\fr{L}{w_+(T_c)},
 \en
and
 \beg \label{alfa}
 \alpha\equiv \fr{L}{4aT_+^4}=\fr{\baL}{3}\fr{T_c^4}{T_+^4}.
 \en
The $+$ sign in Eq. (\ref{steinhardt}) corresponds to weak
detonations or strong deflagrations (i.e., $|v_-|>c_s$), while the
$-$ sign corresponds to strong detonations or weak deflagrations
($|v_-|<c_s$). At the Jouguet point, we have $v_-=-c_{s}$ and
 \beg \label{vj}
 v_+=-v_J\equiv -\frac{1/\sqrt{3}\pm\sqrt{\alpha^2+2
 \alpha/3}}{1+\alpha}.
 \en
Here, the $+$ and $-$ signs correspond to detonations and
deflagrations, respectively.

Using the above relations, we can obtain the stationary wall velocity
from Eq. (\ref{statv}). For deflagrations,  the boundary conditions
give $v_w=-v_-$. Besides, the temperature $T_+$ is obtained by
applying Eqs. (\ref{EM3.1}-\ref{EM3.2}) to the shock front
discontinuity. This gives
\begin{equation}
\frac{\sqrt{3}\left( T _{+}^4-T_N^4\right) }{\sqrt{\left( 3T_+^4+T_N^4\right)
\left( 3T_N^4 + T_+^4\right) }}
=\frac{v_{+}-v_{-}}{1-v_{+}v_{-}}
.  \label{tmatn}
\end{equation}
For detonations, in contrast, the fluid is not reheated in front of
the wall, and we have simpler  conditions, namely, $T_+=T_N$,
$v_w=-v_+$.

The condition (\ref{runaway}) for existence of runaway solutions
gives, for the bag EOS,
 \beg \label{runawaybag}
 1-\fr{T_N^4}{T_c^4}>\fr{4}{\lambda}\fr{\eta}{L}.
 \en
On the other hand, according to Eq. (\ref{nostat}), detonations
cannot exist if
 \beg \label{nostatbag}
 1-\fr{T_-^2T_N^2}{T_c^4}>\fr{4}{\lambda}\fr{\eta}{L}.
 \en

\begin{figure}[bth]
\centering
\epsfysize=5.5cm \leavevmode \epsfbox{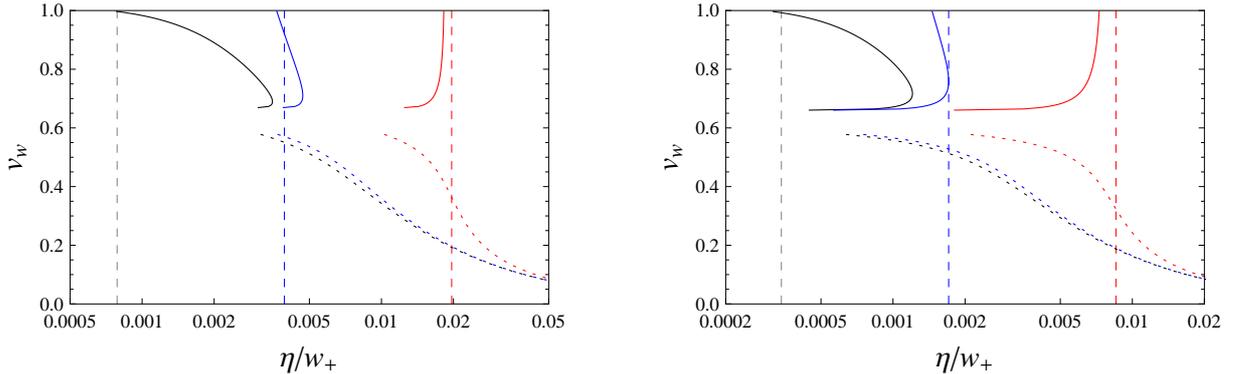}
\caption{The wall velocity as a function of the friction parameter $\eta$, for
$\lambda=0.2$ (black), $\lambda=1$ (blue), and $\lambda=5$ (red). Solid lines correspond
to weak detonations and dotted lines to weak deflagrations. The vertical dashed lines indicate
the values of $\eta$ below which runaway is possible for each value of lambda.
The latent heat parameter is given by $\baL=0.03$, and
the amount of supercooling is $T_N/T_c=0.9$ (left panel), and $T_N/T_c=0.95$ (right panel).}
\label{figvw}
\end{figure}

For a complete description of the solutions, see Ref. \cite{ariel13}.
Here, we are primarily interested in weak detonations. In Fig.
\ref{figvw} we show the weak detonation velocity as a function of the
friction parameter $\eta$ (solid lines). We considered two different
amounts of supercooling (left and right panels), both of which are
strong enough (for the given value of the latent heat) for the
existence of detonations and runaway solutions. For comparison, we
show also the weak deflagration solutions (dotted lines). We
considered a few values of $\lambda$ which give different behaviors
of the friction force, namely, $\lambda=0.2$ (in black), $\lambda=1$
(in blue), and $\lambda=5$ (in red).

It is apparent that the value of $\lambda$ affects especially
supersonic velocities. The value $\lambda=1$ is not special, but is
representative for the case in which the ``friction coefficient''
$|F_{\mathrm{fr}}/v_w|$ has similar values in the non-relativistic
and ultra-relativistic limits. The value $\lambda=0.2$ corresponds to
a high ultra-relativistic friction,
while the value $\lambda=5$ corresponds to a small ultra-relativistic
friction. Except in the limit $\lambda\to 0$, the friction saturates
at high velocities and the detonation approaches the speed of light
at a finite value of $\eta$, which is given by Eq. (\ref{nostatbag}).
Below this value, we have no detonations, and the wall runs away. The
existence of the runaway solution, on the other hand, is determined
by Eq. (\ref{runawaybag}). The value of $\eta$ below which runaway is
already possible is indicated in Fig. \ref{figvw} by the vertical
dashed lines.

For large values of $\eta$, the wall velocity is small and the
hydrodynamic process is a deflagration. As $\eta$ is decreased, the
velocity increases. However, the velocity does not grow as $v_w\sim
1/\eta$, as Eq. (\ref{statv}) may suggest, since hydrodynamic effects
act as an effective friction \cite{ms09,kn11}. Indeed, the reheating
of the plasma slows down the wall, since the driving force
(\ref{fdrbag}) decreases as any of the temperatures $T_\pm$
approaches $T_c$.

\begin{figure}[bth]
\centering
\epsfysize=5cm \leavevmode \epsfbox{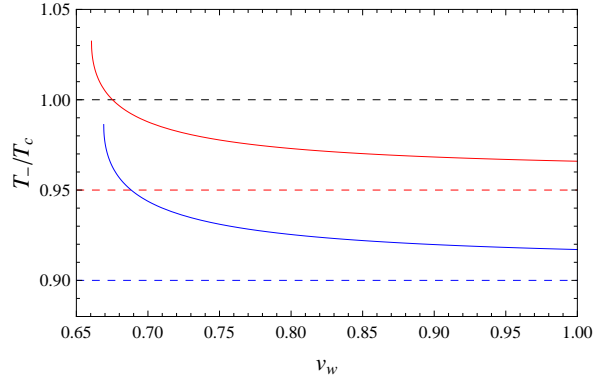}
\caption{The fluid temperature $T_-$ (solid lines) as a function of the detonation
wall velocity, for $\baL=0.03$, and for $T_N/T_c=0.9$ (in blue) and $T_N/T_c=0.95$ (in red).
Dashed lines indicate the values of $T_N$ as well as the critical temperature.}
\label{figreh}
\end{figure}

Below a certain value $\eta=\eta_{\max}$, we have detonations in
addition to deflagrations. Notice that we have in general two weak
detonation solutions. One of them behaves ``normally'' with $\eta$,
i.e., the velocity increases  as the friction decreases. This branch
of solutions ends with a velocity $v_w=1$. The other branch
corresponds to stronger weak detonations and ends at the Jouguet
point. This branch behaves rather unphysically, since the velocity
decreases as the friction decreases. This means that the fluid
disturbances cause a strong friction effect. Indeed, the reheating is
significant near the Jouguet point. This can be appreciated in Fig.
\ref{figreh}, where we show the reheating as a function of $v_w$ for
the two values of $T_N$ considered in Fig. \ref{figvw}. Notice that,
for $v_w\simeq 1$, the temperature $T_-$ is quite close to $T_N$.
However, as the velocity decreases, the reheating increases, and we
may even have $T_->T_c$ as the Jouguet velocity is approached.

It is worth commenting that the deflagration solutions have a similar
behavior, i.e., if the deflagration curve  in Fig. \ref{figvw} is
continued beyond the speed of sound, it becomes double-valued (see,
e.g., \cite{ms12}). However, this occurs for strong deflagrations,
which are unstable \cite{stabdefla}. In the case of detonations, in
contrast, this behavior occurs already for weak solutions, although
the ``unphysical'' branch is closer to the Jouguet point and, thus,
corresponds to stronger weak detonations.

\subsection{Physical and unphysical weak detonations}

The example considered in the previous subsection (the bag EOS) shows
an interesting feature of weak detonations, namely, the existence, in
a certain range of parameters, of two solutions. Although the lower
velocity branch has a strange dependence on the parameters, these
solutions are not physically forbidden. Their behavior is due to
strong hydrodynamics effects. We shall now analyze this feature in
detail, without specifying any equation of state.

Let us consider the dependence of $v_w$ on the friction and the
amount of supercooling. Differentiating Eq. (\ref{statv}), we have
\begin{equation}\label{dif}
- \left\langle{v\gamma_{\lambda}}\right\rangle d\eta + \eta\langle{v^3\gamma^3_{\lambda}}\rangle \lambda d\lambda
 -\fr{\partial F_{\mathrm{dr}}}{\partial T_+} dT_N = Q dv_w,
\end{equation}
with
\begin{equation} \label{Q}
 Q= -
  \fr{\eta}{2}\left(\ga_{\lambda w}^3+\ga_{\lambda-}^3\fr{\pa v_-}{\pa
v_+}\right)
    +
\fr{\pa F_{\mathrm{dr}}}{\pa T_-}
 \fr{T_-}{w_-}\fr{c_{s-}^2}{1+c_{s-}^2}\fr{\pa w_-}{\pa v_w} ,
\end{equation}
where $\pa v_-/\pa v_+$ and ${\pa w_-}/{\pa v_w}$ are given by Eqs.
(\ref{dvmedvma}-\ref{dwmedvma}). Therefore, the derivatives of $v_w$
with respect to any of the parameters $\eta$, $\lambda$, or $T_N$ are
inversely proportional to the factor $Q$, and shear some essential
properties. Consider, for instance, $T_N$ and $\lambda$ fixed. We
have
 \beg \label{detadvw}
\pa \eta/\pa v_w =Q/\langle{-v\gamma_{\lambda}}\rangle.
 \en
Since $-v$ is positive, the sign of $\pa \eta/\pa v_w$ depends on the
two terms in Eq. (\ref{Q}). The first term is negative, since $\pa
v_-/\pa v_+$ is positive for weak detonations. On the other hand, the
second term is positive, since we have ${\pa F_{\mathrm{dr}}}/{\pa
T_-}<0$ (the driving force increases as $T_{\pm}$ decrease) and ${\pa
w_-}/{\pa v_w}<0$, as can be seen in Eq. (\ref{dwmedvma}) (for weak
detonations). The first term in Eq. (\ref{Q}) is due to the
dependence of the friction force on the velocity. This term would
give a ``normal'' variation of $v_w$ with $\eta$ (i.e., $v_w$
decreasing with $\eta$). On the other hand, the second term  comes
from $\pa F_{\mathrm{dr}}/\pa v_w$ and is due entirely to
hydrodynamics. Indeed, notice that the driving force does not depend
on the wall velocity explicitly, but only through the temperature
$T_-$.

The negative sign of ${\pa w_-}/{\pa v_w}$ implies that, the higher
the velocity, the smaller the reheating behind the wall. This means
that the hydrodynamics is always weaker for higher velocities. The
strengthening of hydrodynamics effects as the Jouguet point is
approached is apparent in Eqs. (\ref{dvmedvma}-\ref{dwmedvma}) if we
notice the dependence on the two parameters $v_+-v_-$ and
$v_-^2-c_-^2$. The former is the velocity discontinuity across the
interface, which vanishes for $v_w\to 1$ and is maximum at the
Jouguet point (see Fig. \ref{figvmavme}). In Eqs.
(\ref{dvmedvma}-\ref{dwmedvma}), this maximum is further emphasized
by the divergence due to the factor $v_-^2-c_-^2$ in the
denominators. Thus, for very weak detonations (i.e., in the limit
$v_w\to 1$), we have a finite value of $\pa v_-/\pa v_+$ (namely,
$\pa v_-/\pa v_+= \ga_+^2/\ga_-^2= w_-/w_+$), and a vanishing ${\pa
w_-}/{\pa v_w}$, (hence, a variation of the wall velocity does not
alter the reheating temperature). On the other hand, at the Jouguet
point both $\pa v_-/\pa v_+$ and ${\pa w_-}/{\pa v_w}$ diverge. These
are the behaviors observed for the particular cases considered in
Figs. \ref{figvmavme} and \ref{figreh}.

Obtaining the effect of this hydrodynamics on the wall velocity is
straightforward from Eq. (\ref{detadvw}). For $v_w$ close to $1$, we
will have $\pa v_w/\pa \eta<0$, since the second term in Eq.
(\ref{Q}) will be vanishingly small. On the other hand, at the
Jouguet point we will have $\pa v_w/\pa \eta=0$, since both terms of
$Q$ diverge  (cf. Fig. \ref{figvw} for the particular case of the bag
EOS). Besides, near the Jouguet point, we will see that the second
term in (\ref{Q}) becomes larger than the first one. Hence, we have
$\pa v_w/\pa \eta>0$, i.e., we obtain the ``unphysical branch'' of
weak detonations. Indeed, replacing in Eq. (\ref{Q}) the expressions
for $\pa v_-/\pa v_+$ and ${\pa w_-}/{\pa v_w}$ given by Eqs.
(\ref{dvmedvma}-\ref{dwmedvma}), it can be seen that $Q$ vanishes at
a velocity $v_\mathrm{crit}$ given by the equation
 \beg \label{etamax}
\ga_-^{-1} \left\langle \fr{\ga_\lambda^3}{\ga^2}\right\rangle
\fr{v_-^2-c_{s-}^2}{c_{s-}^2} =
\left(\beta_--\fr{v_-^2+c_{s-}^2}{2c_{s-}^2}\fr{\ga_{\lambda
-}^3}{\ga_-^3}\right)\fr{1-v_+v_-}{1-v_-^2}\fr{\Delta v}{v_+},
 \en
where
 \beg \label{beta}
  \beta_-=\fr{\langle \ga_\lambda v\rangle v_-}{\ga_-}
  \fr{2T_-}{F_{\mathrm{dr}}}\left(-\fr{\pa F_{\mathrm{dr}}}{\pa T_-}\right).
 \en
The velocity  $v_\mathrm{crit}$ corresponds to the maximum
$\eta_{\max}$ of the curve of $\eta$ vs $v_w$. Provided that the
factor in parenthesis in Eq. (\ref{etamax}) is positive, the equation
will be satisfied for a velocity $v_{\mathrm{crit}}$ between the
Jouguet value $v_J$ and the speed of light. We shall see that this is
indeed the case. Hence, the curve of $v_w$ vs $\eta$ is always of the
form of Fig. \ref{figvw}.

To see that the factor in parenthesis in Eq. (\ref{etamax}) is in
general positive, notice that the second term inside the parenthesis
is at most $\sim 1$ for $|v_-|$ between $c_{s-}$ and $1$. On the
other hand, in $\beta_-$ we distinguish two factors of different
nature\footnote{The parameter $\beta_-$ is associated to the fluid
perturbations behind the wall, and will appear again when we consider
the stability of the stationary solution in the next section.}. The
first factor depends on the velocity and is $\sim v_w^2$, whereas the
second factor contains the information on the equation of state. We
may write
 \beg
 \beta_-\sim v_w^2/v_c^2,
 \en
with
 \beg
 v_c^2\propto\fr{F_{\mathrm{dr}}}{{T_-} \pa F_{\mathrm{dr}}/{\pa T_-}}.
 \en
For weak detonations, $v_w$ is supersonic, while the velocity
parameter $v_c$ is in general very small. Indeed, $v_c^2$ gives a
dimensionless measure of the driving force, which is very small for
$T_{\pm}$ close to $T_c$. Hence, we will generally have $\beta_-\gg
1$. In particular, the large reheating near the Jouguet point will
cause a very small $F_{\mathrm{dr}}$. Physically, this effect
prevents the velocity to increase as the friction decreases, and
causes the lower velocity branch of weak detonations.

For the sake of concreteness, let us consider again the bag EOS, for
which we have
 \beg \label{betabag}
 \beta_-=\fr{\gamma_-^{-1}\langle\gamma_\lambda v\rangle
 v_-}{\fr{1}{4}\left(\fr{T_c^4}{T_-^2T_+^2}-1\right)}\hspace{.5cm} \textrm{(bag)}.
 \en
For small reheating ($T_-\simeq T_+= T_N$), the denominator is given
by $ {\fr{1}{4}\left({T_c^4}/{T_N^4}-1\right)}$. Consider for
instance the electroweak phase transition. For different extensions
of the Standard Model, the value of $T_N/T_c$ is typically in the
range $0.8-1$, while very strong phase transitions may reach a
supercooling of $T_N/T_c\simeq 0.7$ \cite{stabdefla}. For this lower
limit, we have ${\fr{1}{4} \left({T_c^4}/{T_N^4}-1\right)}\simeq 0.8
$. However, a large supercooling will be accompanied in general by a
large latent heat, since both are characteristics of a strong phase
transition. Even disregarding the reheating caused by the release of
latent heat, for such a large supercooling we expect a high velocity,
$v_w> 0.9$, and the numerator in (\ref{betabag}) will be $\sim
v_w^2>0.8$. Therefore, even in such an extreme case, we expect
$\beta_-\gtrsim 1$. Due to reheating, the denominator in Eq.
(\ref{betabag}) will be in general quite smaller than ${\fr{1}{4}
\left({T_c^4}/{T_N^4}-1\right)}$. As we have seen, we may even have
$T_->T_c$. Thus, in the general case we will have $\beta_-\gg 1$.

\section{Stability analysis of weak detonations} \label{staban}

The structure of multiple stationary solutions observed in Fig.
\ref{figvw} was found in different numerical calculations (see e.g.
\cite{ikkl94}). There exist also supersonic Jouguet deflagrations
(not shown in Fig. \ref{figvw} for simplicity), which also coexist
with detonations in some parameter ranges. Besides, as we have seen,
there are ranges of coexistence with runaway walls. The coexistence
of hydrodynamic solutions is an important issue, since one has to
determine which of these will be realized during the phase
transition. Investigating the stability of the stationary motion may
help deciding which solutions to choose. In particular, one expects
that the ``unphysical'' weak detonations described above should not
be realized \cite{ms09}. As a mater of fact, numerical simulations
suggest that the detonations which are closer to the Jouguet point
are unstable \cite{ikkl94}. We shall now consider the linear
stability of the weak detonation solution.

\subsection{Linear perturbations}

We shall consider small variations of the wall position $\z(x^\pe,t)$
around a planar interface. For planar symmetry, we only need to
consider a single direction $x^\pe$ transverse to the wall motion.
Besides, we need to consider, on each side of the wall, the
longitudinal perturbation of the velocity $\de v (x^\pe,z,t)$ and the
transverse velocity $v^\pe(x^\pe,z,t)$, as well as the pressure
fluctuation $\de p(x^\pe,z,t)$ (the latter may be replaced by the
temperature fluctuation $\delta T$).

The standard approach  (see, e.g.,
\cite{landau,stabdefla,link,hkllm}) consists in considering small
perturbations around a solution with $v=\mathrm{const}$ and
$T=\mathrm{const}$. This allows to consider Fourier modes and, hence,
to deal with algebraic equations (dispersion relations). It is worth
remarking that this assumption is not always valid. Such is the case,
for instance, of the fluid profile for the Jouguet detonation (Fig.
\ref{figdeto}, right panel). Perturbing this profile  gives much more
involved equations and is out of the scope of this work.

For the weak detonation solution depicted in the left panel of Fig.
\ref{figdeto}, the fluid profile is a constant in front of the wall.
Behind the wall, the profile  is a constant up to a point $z_0=\xi_0
t$, with $\xi_0$ given by Eq. (\ref{xi0}). Beyond this point, we have
the rarefaction wave $v_{\mathrm{rar}}(z/t)$. In general, assuming a
constant profile is a good approximation for perturbations originated
at the wall. Indeed, the Fourier modes will decay as $e^{-q|z|}$ away
from the wall, within a distance scale $q^{-1}$ which will be given,
in general, by $q^{-1}\sim d$. In contrast, the zone of constant $v$
and $T$ will grow quickly to values $|\xi_0 t|\gg d$, since the scale
$d$ is much smaller than the time associated to bubble expansion [see
the discussion below Eq. (\ref{d})]. This approach will break down,
anyway, for weak detonations which are very close to the Jouguet
point, where $\xi_0$ vanishes.

\subsubsection{Fluid equations and junction conditions}

The equations for the fluid perturbations have been considered
recently in Ref. \cite{stabdefla}, and we shall only write down the
results. We shall consider a reference frame moving with the
unperturbed wall. The equations for the fluid perturbations, as well
as their matching conditions at the interface, are derived by
considering Eqs. (\ref{fluideq})  as in the previous section, this
time  for the perturbed variables, and keeping to linear order in the
perturbations. For the fluid away from the wall, we have
 \bega\label{HD6}
    c_s^2 {w}({\ga}^2{v}\de v{,_0}+{\ga}^2\de v_{,z}+v^\pe_{,\pe})+\de p_{,0}+{v}\de
    p_{,z}=0, \\
  {w}{\ga} ^2(\de v_{,0}+{v}\de v_{,z})+{v}\de p_{,0}+\de p_{,z}=0, \label{HD7} \\
  {w} {\ga}^2(v^\pe_{,0}+{v}v^\pe_{,z})+\de p_{,\pe}=0, \label{HD8}
 \ena
while the matching conditions at the interface are given by
 \bega
    \Delta\left[\baw\baga^2(1+\bav^2)(-\pa_0\z+\baga^2\de v)+
    (1+c_s^{-2})\baga^2\bav\de p\right]=0,\label{junc1}\\
    \Delta(v^\pe+\bav\pa_\pe\z)=0,\label{junc2}\\
    \sigma(\pa_0^2-\pa_\pe^2)\z+\Delta\left[2\baw\baga^4\bav\de v+
    \left(1+(1+c_s^{-2})\baga^2\bav^2\right)\de p\right]=0,
    \label{junc3}
 \ena
where $\Delta$ applied to any function $f$ means $f_+-f_-$. Notice
that these equations depend on the equation of state only through the
parameter $c_s$.

\subsubsection{Equation for the interface}

The equation for the interface perturbations is obtained from  the
field equation (\ref{fieldeq}), like in the previous section. We have
already derived this equation in Ref. \cite{stabdefla}. However, we
considered a simplified version of the damping term in Eq.
(\ref{fieldeq}), corresponding to the particular case $\lambda=0$. As
can be seen in Fig. \ref{figvw}, this is generally a good
approximation for deflagrations but not for detonations, which depend
strongly on $\lambda$. We shall consider here the general case.

We consider  perturbations around  a stationary wall, which is at
$z_w=0$, and for which we have $\dot z_w=\ddot z_w=0$. On the other
hand, the perturbed wall is  at the position
$z_w=\zeta(x^{\perp},t)$. We thus assume a field profile\footnote{The
small perturbation $\zeta$ is macroscopic in comparison with the wall
width $l_w$.} of the form
$\phi(z,x^{\perp},t)=\phi[z-\z(x^{\perp},t)]$. To first order in $\z$
and $v^{\pe}$, we have
$\pa_{\mu}\pa^{\mu}\phi=\phi'(\pa_{\pe}^2-\pa_0^2)\z-\phi''$ and
$u^{\mu}\pa_{\mu}\phi=\gamma(-\pa_0\zeta+v)\phi'$. Multiplying  Eq.
(\ref{fieldeq}) by $\phi'(z-\z)$ and integrating in $z$, we obtain
 \beg\label{EM5}
    \si(\pa_0^2-\pa_\pe^2)\z=
    F_{\mathrm{dr}}[T]+ F_{\mathrm{fr}}[\gamma(v-\pa_0\z)],
 \en
where the forces $F_{\mathrm{dr}}[T]$, $F_{\mathrm{fr}}[\gamma v]$
are  given by Eqs. (\ref{fdrexac}) and (\ref{feno}) as functionals of
the temperature and velocity configurations. We notice the following
differences with Eq. (\ref{fuerzas}). The term $-\si\pa_\pe^2\z$
gives the restoring force due to the curvature of the surface. The
argument of the driving force is now the perturbed temperature, $T\to
T+\delta T$. The velocity $v$ in the friction force is the perturbed
one, $v\to v+\delta v$, and the argument of $F_{\mathrm{fr}}$ is
further modified by the term $-\pa_0\z$. To understand this
dependence, notice that the friction must depend on the
\emph{relative} velocity $v_r$ between the fluid and the wall, which
is given by the relation $\ga_r v_r=\gamma(v-\pa_0\z)$.

We may use the approximations for the force functionals derived in
Sec. \ref{stationary} in terms of the values of the variables outside
the wall,
 \beg
F_{\mathrm{dr}}[T]\simeq F_{\mathrm{dr}}(T_+,T_-), \hspace{.5cm}
F_{\mathrm{fr}}[\gamma v]\simeq
F_{\mathrm{fr}}(\gamma_+v_+,\gamma_-v_-),
 \en
where we must do the replacements $T\to T+\delta T$, $\gamma v\to
\gamma v+\delta(\gamma_r v_r)$ in each argument. Hence, the
perturbations from the stationary case are given by
 \bega
  \delta F_{\mathrm{dr}}&=&\fr{\pa F_{\mathrm{dr}}}{\pa T_+}\delta T_+
    +\fr{\pa F_{\mathrm{dr}}}{\pa T_-}\delta T_-, \\
  \delta F_{\mathrm{fr}} &=& \fr{\pa F_{\mathrm{fr}}}{\pa (\ga_+v_+)}\left[\de(\ga_+
  v_+)-\ga_+\pa_0\z\right]+\fr{\pa F_{\mathrm{fr}}}{\pa
  (\ga_-v_-)}\left[\de(\ga_-
  v_-)-\ga_-\pa_0\z\right].
 \ena
Using Eq. (\ref{fric}), Eq. (\ref{EM5}) gives
 \beg\label{EM9}
    \si(\pa_0^2-\pa_\pe^2)\z=
    2 \left\lan \fr{\pa F_{\mathrm{dr}}}{\pa T}\delta T\right\ran +
    \eta \left\lan \fr{\de(\ga
    v)-\baga\pa_0\z}{(1+\lambda^2\gamma^2v^2)^{3/2}}\right\ran.
 \en
The parameters $\sigma$ and $\eta$ can be written in terms of the
fluid velocity and the scale $d$ using Eqs. (\ref{statv}) and
(\ref{d}). Furthermore, the temperature variation is related to our
perturbation variable $\delta p$ through $\delta p=s\delta T$. Thus,
we may write
 \beg \label{interf}
 \left\langle \ga_\lambda v \,d \, (\pa_0^2-\pa_\pe^2)\z + \ga_\lambda^3\de
  v -   \fr{\ga_\lambda^3}{\ga^2}\pa_0\z +\fr{\gamma \beta}{v}\fr{\de p}{w}\right\rangle=0,
 \en
where the parameters $\beta_\pm$ contain information about the equation of
state (we have already introduced $\beta_-$ in the previous section),
 \beg
 \label{betapm2}
  {\beta_{\pm}}= \fr{\lan \baga_\lambda\bav\ran v_\pm}{\ga_\pm}\fr{2T_{\pm}}{F_{\mathrm{dr}}}
  \left(-\fr{\pa F_{\mathrm{dr}}}{\pa T_{\pm}}\right) =
  \fr{\lan \baga_\lambda\bav\ran v_\pm}{\ga_\pm} \fr{4T^2_{\pm}}{F_{\mathrm{dr}}}
  \left(-\fr{\pa F_{\mathrm{dr}}}{\pa
  T^2_{\pm}}\right).
 \en
The last equality is useful if $F_{\mathrm{dr}}$ is quadratic in the
temperature \cite{stabdefla}.

\subsection{Fourier modes of the perturbations} \label{fourier}

The fluid equations away from the wall, Eqs. (\ref{HD6}-\ref{HD8}), can be
expressed in  matrix form, defining the perturbation vector
 \beg \label{U}
  {\vec{U}}\equiv\left[\begin{array}{c}
                      \de p\\
              \de v\\
              v^\pe
                     \end{array}\right].
 \en
We are interested in solutions of the form
 \beg\label{formaU}
    {\vec{U}}(t,z,x^\pe)=\vec{L}e^{\Omega t+qz+ikx^{\pe}},
 \en
where  $k$ is a real wavenumber corresponding to Fourier modes along
the wall, and $\Omega,q$ are in general complex numbers.  The
stationary solution will be unstable if there are modes with
$\mathrm{Re} (\Omega)>0$. Inserting Eqs. (\ref{U}-\ref{formaU}) into
the fluid equations we obtain an eigenvalue equation for the
perturbation modes (for details, see \cite{stabdefla}). This gives
the dispersion relations
 \bega
    {q}_1&=&-{\Omega}/{v} ,\label{Fou10}\\
  q_{2,3}&=&\fr{ (1-c_s^2)v\Omega \pm c_s(1-v^2)\sqrt{\Omega^2+
   ({c_s^2-v^2})\ga^2 k^2}}{c_s^2-v^2}, \label{Fou11}
 \ena
corresponding to the eigenvectors
 \beg\label{Fou7}
      \vec{L}_1=\left[\begin{array}{c}
                      0\\
              1\\
              \fr{iq_1}{k}
                     \end{array} \right],\;
      \vec{L}_{2,3}=\left[\begin{array}{c}
                      -w\ga^2\left(\fr{\Omega+q_{2,3}v}{\Omega v+q_{2,3}} \right)\\
              1\\
              \fr{ik}{\Omega v+q_{2,3}}
                     \end{array} \right]
 \en
The eigenvector $\vec{L}_1$ is a special mode corresponding to a
perturbation with $\de p=0$, which moves with the fluid (i.e., the
perturbation is a function of $z-v t$).

\begin{figure}[bth] \centering \epsfysize=5.5cm \leavevmode
\epsfbox{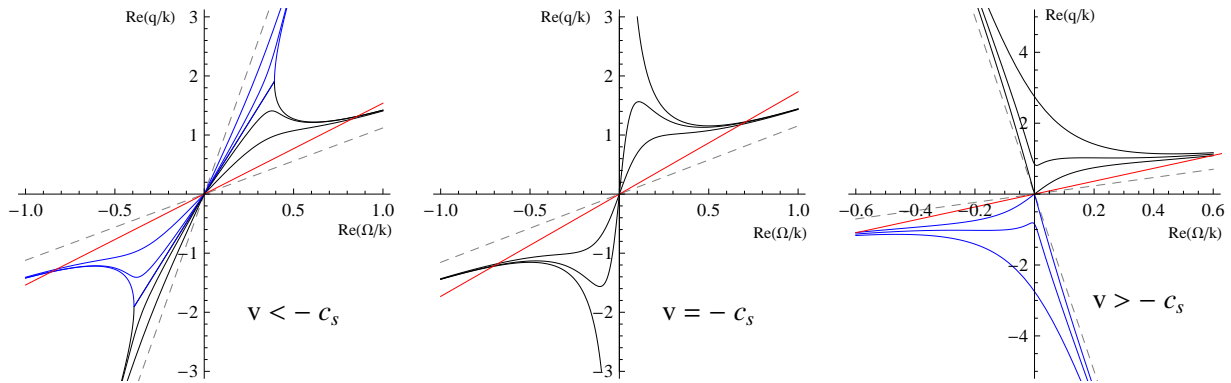} \caption{The real parts of the dispersion
relations $q_1(\Omega)$ (in red), $q_2(\Omega)$ (in black), and $q_3(\Omega)$
(in blue), for
different values of the imaginary part $\mathrm{Im}(\Omega)$. Gray
dashed lines indicate the asymptotes of these solutions.}
\label{fighip}
\end{figure}

In Fig. \ref{fighip} we show the real part of $q$ as a function of
the real part of $\Omega$ for the three modes. For the special mode
we have a linear relation between $q$ and $\Omega$, $q_1=\Omega/|v|$.
Regarding the solutions $q_{2,3}$, the relation is linear for $k=0$,
 \beg \label{asint}
  q_{2,3}=\Omega/a_{2,3},
 \en
with
 \beg
 a_{2,3}=\fr{|v|\pm c_s}{1\pm c_s |v|}.
 \en
For $k\neq 0$, the lines (\ref{asint}) are asymptotes for the curves
of $\mathrm{Re}(q)$ vs. $\mathrm{Re}(\Omega)$. These asymptotes
depend on the value of $v$.
We are interested in the weak detonation case, for which both $v_+$
and $v_-$ are supersonic (left panel of Fig. \ref{fighip}). In this
case, $\mathrm{Re}(q)$ and $\mathrm{Re}(\Omega)$  have always the
same sign. Notice that, in the limit $|v|=c_s$, one of the asymptotes
becomes vertical and we are left with only two modes (central panel);
specifically, we have $q_1=-\Omega/v$ and
 \beg\label{Fou6}
    q_2=\fr{c_s k^2}{2\Omega}+\fr{(1+c_s^2)}{2c_s}\Omega.
 \en
The subsonic case (right panel) is only relevant for deflagrations or
strong detonations.

For given $k$ and $\Omega$, the general solution will be a superposition of
these modes,
 \beg
    {\vec{U}}(t,z,x^\pe)=\vec{A}(z)e^{(\Omega t+ikx^\pe)},
 \en
with
 \beg\label{vecA}
    \vec{A}(z)=\sum_{j=1}^3 A_j \vec{L}_je^{q_j z},
 \en
where the constants $A_j$ are different on each side of the wall. The
function $\vec{A}(z)$ must satisfy the junction conditions at the wall. On
the other hand, the surface perturbation will be of the form
 \beg\label{D}
        \z(t,x^\pe)=D e^{(\Omega t+ikx^\pe )}.
 \en
Notice that, for weak detonations, we always have  $\mathrm{Re}(q)=0$ for
$\mathrm{Re}(\Omega)=0$. This case corresponds to undamped oscillations of
the wall which generate plane waves in the fluid. Since we are looking for
instabilities, we are interested in the case $\mathrm{Re}(\Omega)>0$. In
this case, we have always $\mathrm{Re}(q)> 0$. The condition that the source
of instabilities is the wall itself, and not something outside it, implies
that the perturbations must decrease with the distance from the wall
\cite{landau}. Therefore, we must require $\mathrm{Re} (q)<0$ for $z>0$
(i.e, in the $+$ phase) and $\mathrm{Re} (q)>0$ for $z<0$ (i.e., in the $-$
phase).

The requirement $\mathrm{Re}(q)<0$ in front of the wall  implies, for
detonations, $\mathrm{Re}(\Omega) < 0$. This means that perturbations of the
fluid in front of the wall decay exponentially. This is a consequence of the
fact that the fluid enters the wall supersonically. The absence of unstable
modes in the $+$ phase led the authors of Ref. \cite{hkllm} to the
conclusion that detonations are always stable. However, as pointed out in
Ref. \cite{abney}, perturbations in the $-$ phase should also be considered.
Behind the wall we require $\mathrm{Re}(q)>0$, for which we have
$\mathrm{Re}(\Omega)> 0$ in the weak detonation case. Such a perturbation
will grow exponentially. Thus, weak detonations may be unstable.
Nevertheless, not any fluid perturbation will fulfill the junction
conditions. We have just seen that perturbations which are localized near
the wall have a different time dependence in front and behind the wall. Such
perturbations cannot be matched. Therefore, if we look for unstable
perturbations, we must consider a solution which is of the general form
(\ref{vecA}) behind the wall and  \emph{vanishes} in front it.

The approach of previous works  \cite{abney,r96} to the stability of
detonation fronts in a cosmological phase transition consisted essentially
in solving the aforementioned matching conditions. However, the general
treatment of these works is more suitable for chemical burning than for a
phase transition. Although the case of chemical burning is out of the scope
of the present paper, we wish to point out that, even in that case, some
aspects of the treatment of Refs. \cite{abney,r96}  seem to be incorrect.

In the first place, only strong and Jouguet detonations were
considered in Refs. \cite{abney,r96}. We remark that, for a phase
transition front, strong detonations are impossible, since the
stationary profile cannot satisfy the boundary and junction
conditions (see, e.g., \cite{s82,l94}). Therefore, this solution
should not be considered from the beginning. In the case of chemical
burning, according to the Chapman-Jouguet theory, strong detonations
should decay to a Jouguet detonation.

In contrast, in the case of chemical burning, weak detonations are not
possible. According to an argument by Steinhardt \cite{s82}, this would also
be the case of a cosmological phase transition. However, as already pointed
out by Landau \cite{landau}, a phase transition front must be treated
differently than a burning front. In Ref. \cite{l94} it was shown, in
particular, that weak detonations are possible in a cosmological phase
transition.

In this regard, we may note, for instance, that in the case of
chemical burning, the reheating of the fluid favors the combustion
and, thus, the propagation of the burning front. In contrast, in the
case of a phase transition, the reheating causes a decrease of the
pressure difference between phases and, thus, opposes to the
propagation of the front.

This last feature is taken into account by the interface equation,
either Eq. (\ref{statv}) for the stationary case or Eq.
(\ref{interf}) for the perturbed wall. No equation for the interface
was considered in Refs. \cite{abney,r96}. Therefore, only the three
junction conditions (\ref{junc1}-\ref{junc3}) were imposed on the
system. In the case of Jouguet or strong detonations, this is
compatible with the number of unknowns. Indeed, notice that, in the
Jouguet case (central panel of Fig. \ref{fighip}), we have only two
unstable modes in the $-$ phase. Similarly, it can be seen from the
right panel of Fig. \ref{fighip} that, for strong detonations, we
also have only two unstable modes. We thus have three unknowns,
namely, the coefficients $A_1$ and $A_2$ of these modes, plus the
coefficient $D$ of the surface perturbation. By solving this system
of equations, it was found in Ref. \cite{abney} that unstable modes
exist at all wavelengths. In contrast with this result, in Ref.
\cite{r96} it was found, with a similar treatment,  that strong
detonations can be stable and that Jouguet detonations are
unconditionally stable.

In any case, an important feature seems to have been missed by the authors
of  Refs.  \cite{abney} and \cite{r96}, namely, that the  usual approach of
considering perturbations around a constant solution will break down for the
profile of a Jouguet detonation, as already discussed.

Let us go back to the phase transition case. Notice that, if we take
into account the surface equation (\ref{interf}) together with the
junction conditions (\ref{junc1}-\ref{junc3}), then we have, for the
Jouguet and the strong detonation, four equations for the three
unknowns $A_1,A_2,D$. Therefore, these cases are not evolutionary,
and the linear perturbation theory breaks down. This is just  a sign
of the problems discussed above for these solutions. In contrast, as
seen in the left panel of Fig. \ref{fighip}, in the weak detonation
case we have three unstable modes behind the wall. Therefore, there
are four unknowns, namely, $A_1,A_2,A_3$, and $D$, and the weak
detonation is evolutionary.

\subsection{Solving the perturbation equations} \label{solu}

Let us thus consider a solution of the form $ \vec A_+ (z)=0$  for
$z>0$, and
 \beg
\vec{A}_-(z)=A_1\vec{L}_{1}e^{{q}_{1} z}+
A_2\vec{L}_{2}e^{{q}_{2}z}+A_3\vec{L}_{3}e^{{q}_{3}z}
\label{modosstrong}
 \en
for $z<0$. The junction conditions (\ref{junc1}-\ref{junc3}), as well
as the surface equation (\ref{interf}), require evaluating Eq.
(\ref{modosstrong}) at the interface position $z=\z(x^{\pe},t)$.
However, to first order in the perturbations, this is equivalent to
evaluating at $z=0$. Therefore we have, in front of the wall,
 \beg\label{Weak2}
  \de v_+= 0,\;\;
  \de p_+=0,\;\;
    v^\pe_+=0,
 \en
and behind the wall (omitting a factor $e^{\Omega t+ik x^\pe}$),
 \bega\label{Weak4}
\de v_-&=&A_1+A_2+A_3,\\ \de p_- &=& -w_-\ga_-^2\left[\fr{\Omega+{
q}_{2}v_-}{\Omega v_- +{q}_{2}}A_2 +
\fr{\Omega+{q}_{3}v_-}{\Omega v_- +{q}_{3}}A_3 \right], \\
v^\pe_-&=&\fr{i{q}_{1}}{k} A_1 + \fr{ik}{\Omega v_- +{q}_{2}}A_2 +
\fr{ik}{\Omega v_- +{q}_{3}}A_3,
 \ena
On the other hand, for the perturbation of the wall we have (omitting
again a factor $e^{\Omega t+ik x^\pe}$)
 \beg
\pa_0\z=\Omega D, \; \pa_{\perp}\z=ikD , \; \pa_0^2\z=\Omega^2 D, \;
\pa_{\perp}^2\z=-k^2D. \label{corrug}
 \en
Inserting Eqs. (\ref{Weak2}-\ref{corrug}) in Eqs.
(\ref{junc1}-\ref{junc3},\ref{interf}), we obtain a homogeneous
system of linear equations for the constants $A_1,A_2,A_3$ and $D$.
It is convenient to redefine the unknowns as $\tilde
A=(\ga_-^2/\ga_{s-}^2)A_3/{Q}_3$, $\tilde
B=(\ga_-^2/\ga_{s-}^2)A_2/{Q}_2$, $\tilde C=-\ga_-^2 A_1$, $\tilde
D=kD$, where the quantities $\gamma_{s-}^2$ and $Q_{2,3}$ are defined
below. After some manipulations, the system of equations can be
written in matrix form as (from now on, we use the notation $k=|k|$)
 \begin{equation}
 \label{matriz}
  \left[
\begin{array}{cccc}
 - \frac{1}{R} &  \frac{1}{R} & \frac{\hat\Omega}{v_-\gamma_-^2} & -{\Delta v} \\
  -(1+\frac{v_-\W}{R}) & -(1-\frac{v_-\W}{R}) & 1+v_-^2 & (1-v_-v_+)\fr{\Delta v}{v_+} \W  \\
 -(1+\frac{\W}{v_-R}) & -(1-\frac{\W}{v_-R}) & 2 & \frac{F_{\mathrm{dr}}}{w_+v_+\ga_+^2}{(\W^2+1)kd} \\
  \fr{\ga_{\lambda-}^3}{\ga_-^3}Q_3-\fr{\beta_-}{v_-}P_3 &
  \fr{\ga_{\lambda-}^3}{\ga_-^3}Q_2-\fr{\beta_-}{v_-}P_2 &
  \fr{-\ga_{\lambda-}^3}{\ga_-^3\ga_{s-}^2} &
    \fr{2}{\ga_{s-}^2}{\left[\fr{\langle\ga_\lambda v\rangle}{\ga_-}(\W^2+1)kd
    -\left\langle\fr{\ga_{\lambda}^3}{\ga^2}\right\rangle\fr{\W}{\ga_-} \right]}
\end{array}%
\right]
 \left[ \begin{array}{c}
  \tilde A \\ \tilde B \\ \tilde C \\ \tilde D
 \end{array}\right]
 =0,
  \end{equation}
where we have defined the quantities $\W= {\Omega}/{k}$, and
 \beg
 \fr{1}{\gamma_{s-}^{2}}={1-\fr{v_-^2}{c_{s-}^2}}, \;\;
 R=\sqrt{\frac{\gamma_-^2}{\gamma_{s-}^2}+\frac{\hat{\Omega}^2}{c_{s-}^2}},\;\;
P_{\substack{2\\ 3}}=v_- \pm \frac{\W}{R}, \;\; Q_{\substack{2\\
3}}=1\pm \frac{v_-}{c^2_{s-}}\frac{\W}{R}.
 \en
(notice that, for weak detonations, we have $\ga_{s-}^2<0$). Nontrivial
solutions exist if the determinant of the matrix in Eq. (\ref{matriz})
vanishes.

Due to the symmetry of the matrix, the $4\times 4$ determinant is easy to
calculate. Indeed, calling $\det_{ij}$ the determinant of the $3\times 3$
matrix that results by removing the $i$-th row and the $j$-th column, we
find that $\det_{14}=0$, and we have the equation
 \beg
   \left[\fr{F_{\mathrm{dr}}}{w_+}\fr{\det{}_{34}}{v_+\ga_+^2}-\fr{2\langle\ga_\lambda
    v\rangle\det{}_{44}}{\ga_{s-}^2}\right]kd(\W^2+1)-\left[\fr{\Delta v}{v_+}(1-v_+v_-)\det{}_{24}-\left\langle
    \fr{\ga_\lambda^3}{\ga^2}\right\rangle\fr{2\det{}_{44}}{\ga_{s-}^2} \right]\W=0. \label{kd}
 \en
Moreover, all of the $3\times 3$ determinants are proportional to
 \beg \label{factor}
 \fr{2}{R}\left(1-\fr{\W^2}{v_-^2\ga_-^2}\right)
 \en
Therefore, a solution for $\Omega>0$ is given by $\W=-\gamma_-v_-$.
For this value of $\W$, we have $R=\ga_-$, $P_2=0$,
$Q_2=\gamma_{s-}^{-2}$, and we obtain, from Eq. (\ref{matriz}),
$\tilde B-\tilde C=\tilde A=\tilde D=0$. This gives
$A_1+A_2=A_3=D=0$. As a consequence, all the perturbation variables
in (\ref{Weak4}-\ref{corrug}) vanish in this case (notice that
$\Omega+q_2v_-\propto P_2, \Omega v_-+q_2\propto Q_2$). Hence, this
is a spurious solution, like in the case of deflagrations
\cite{hkllm,stabdefla}. Omitting the factor (\ref{factor}), which
cancels out in Eq. (\ref{kd}), the determinants are given by
 \beg
 \det{}_{24}= \fr{\ga_{\lambda-}^3}{\ga_-^3}\fr{c_{s-}^2+v_-^2}{c_{s-}^2} -2\beta_-,
 \;\;
 \det{}_{34}= \fr{\ga_{\lambda-}^3}{\ga_-^3}v_-^2\fr{c_{s-}^2+1}{c_{s-}^2} -\be_-(1+v_-^2),
 \;\;
 \det{}_{44}= \fr{1}{\ga_-^2}.
 \en
Notice that $\W$ does not appear in these expressions. Hence, Eq.
(\ref{kd}) gives a simple quadratic equation for $\Omega$. The
solution is
 \beg
 \Omega d=\mathcal{C}\pm\sqrt{\mathcal{C}^2-(kd)^2}, \label{Omega}
 \en
where the real constant $\mathcal{C}$ is given by
 \beg \label{C}
 \mathcal{C}=\fr{1}{2|v_-|}\fr{\mathcal{N}}{\mathcal{D}},
 \en
with
 \bega \label{B}
 \mathcal{N}&=&
 -\fr{1}{\ga_-}\left\langle\fr{\ga_\lambda^3}{\ga^2}\right\rangle\fr{v_-^2-c_{s-}^2}{c_{s-}^2}
 +\fr{1-v_+v_-}{1-v_-^2}\left[\beta_--
            \fr{\ga_{\lambda-}^3}{\ga_-^3}\fr{c_{s-}^2+v_-^2}{2c_{s-}^2}\right]\fr{\Delta v}{v_+},
 \\
 \mathcal{D}&=&\fr{\langle\ga_\lambda v\rangle}{\ga_-
 v_-}\fr{v_-^2-c_{s-}^2}{c_{s-}^2}-
   \fr{1+v_-^2}{2v_-^2} \left[\beta_--\fr{\ga_{\lambda -}^3}{\ga_-^3}
   \fr{1+c_{s-}^2}{1+v_-^2}\fr{v_-^2}{c_{s-}^2}
   \right]\fr{F_{\mathrm{dr}}}{w_-}. \label{A}
 \ena

\begin{figure}[bth] \centering \epsfysize=5cm \leavevmode
\epsfbox{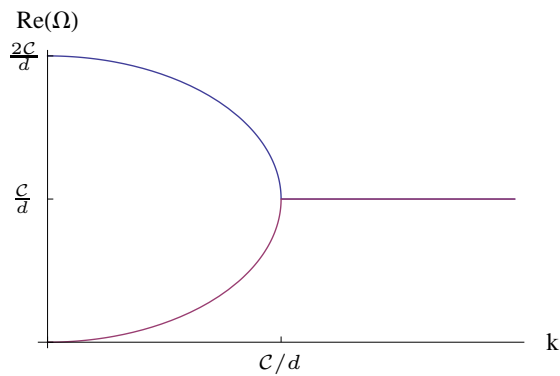} \caption{The real part of $\Omega$
as a function of $k$, for $\mathcal{C}>0$.}
\label{figomegak}
\end{figure}

The structure of the solutions $\Omega(k)$ is very simple. For
$kd<|\mathcal{C}|$ we have two real solutions, whereas for
$kd>|\mathcal{C}|$ we have two complex solutions with the same real
part. In either case, we have $\mathrm{Re}(\Omega)>0$ if, and only
if, $\mathcal{C}>0$ (see Fig. \ref{figomegak}). If this is the case,
perturbations at all wavenumbers are unstable. As we discuss below,
the constant $\mathcal{C}$ will take positive as well as negative
values in different velocity intervals. For $\mathcal{C}<0$, we have
$\mathrm{Re}(\Omega)<0$ for all $k$. It is important to remember,
though, that we are considering perturbations behind the wall, which,
for $\mathrm{Re}(\Omega) <0$ increase exponentially with the distance
from the wall [i.e., correspond to $\mathrm{Re}(q)<0$] and must be
discarded. Since there is also no solution with $\mathrm{Re}(\Omega)
>0$, in this case there must be solutions corresponding to undamped
oscillations. Studying these solutions, which correspond to
$\mathrm{Re}(\Omega)=\mathrm{Re}(q)=0 $, is not one of the goals of the
present paper. In any case, such solutions with $\mathrm{Re}(\Omega)=0$ are
marginally stable, and, to determine the stability, one should go beyond
linear perturbations.

For completeness, we have also considered perturbations in front of
the wall, for which the treatment is very similar. As we have seen,
in this case, the solutions which must be discarded are those with
$\mathrm{Re}(\Omega)>0$. We found again two solutions of the form
(\ref{Omega}). Nevertheless, in this case, it can be seen that the
stability parameter is always negative. This indicates that any
perturbation in front of the wall is exponentially stable, as
expected physically.

Before going on to the determination of the instability intervals, we
wish to compare these results with the deflagration case
\cite{hkllm,stabdefla}. In brief, deflagrations with velocities above
a critical velocity $v_c$ are stable under perturbations at any
wavelength. Below the velocity $v_c$, there is a range of unstable
wavenumbers, $0<k<k_c$. Thus, short wavelengths, as well as long
wavelengths, are stable. In contrast, for detonations all wavenumbers
are either stable or unstable, depending on the wall velocity. It is
interesting that the critical velocity $v_c$ arises due to the
dependence of $\Omega$ on terms which are roughly of the form
$\beta_\pm-1$. A similar dependence is present in Eqs.
(\ref{B}-\ref{A}). Indeed, except for the limit $v_w\to 1$, most of
the factors in Eqs. (\ref{B}-\ref{A}) are $\sim 1$, and we may write,
roughly,
 \bega \label{Bapp}
 \mathcal{N}&\sim&
 -\fr{v_-^2-c_{s-}^2}{c_{s-}^2}
 +\left(\beta_--
            1\right)\fr{\Delta v}{v_+},
 \\
 \mathcal{D}&\sim&\fr{v_-^2-c_{s-}^2}{c_{s-}^2}-
   \left(\beta_--1
   \right)\fr{F_{\mathrm{dr}}}{w_-}. \label{Aapp}
 \ena
In the deflagration case, the parameters $\beta_\pm$ are critical for
stability, since, as we have seen in the previous section, we have
roughly $\be_\pm \sim v_w^2/v_c^2$. Since $v_c$ is generally small,
for a deflagration the factors $\beta_\pm-1$ will change sign at
$v_w\simeq v_c$. In contrast, for weak detonations we generally have
$\beta_-\gg 1$. In this case, the sign of the numerator $\mathcal{N}$
will depend essentially on the balance between two parameters
characterizing the hydrodynamics, namely, $(v_-^2-c_{s-}^2)$ and
$\Delta v$. On the other hand, the sign of the denominator
$\mathcal{D}$ depends on $(v_-^2-c_{s-}^2)$ and $F_{\mathrm{dr}}$.

\section{Stability of weak detonations} \label{stab}

\begin{figure}[bt] \centering \epsfysize=5cm \leavevmode
\epsfbox{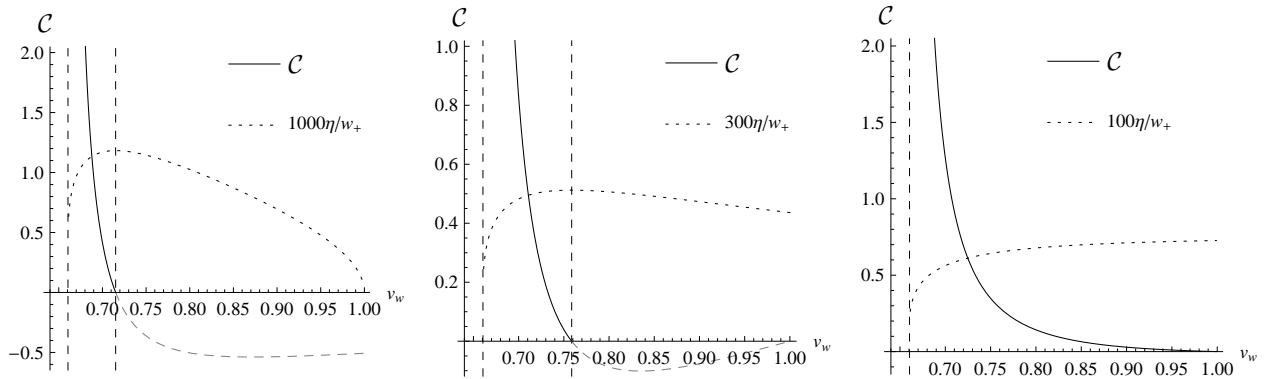} \caption{The stability parameter $\mathcal{C}$  as a function of $v_w$, for the bag model with
$\baL=0.03$, $T_N/T_c=0.95$, and for $\lambda=0$ (left), $\lambda=1$ (center), and $\lambda=5$ (right).
The vertical lines indicate the values of $v_J$ and $v_{\mathrm{crit}}$.}
\label{figstab}
\end{figure}

Let us first consider a specific example.  In Fig. \ref{figstab} we
plot the value of  the stability parameter $\mathcal{C}$ as a
function of $v_w$, for the cases considered in the right panel of
Fig. \ref{figvw}. For the case of small $\lambda$, we have changed
the value $\lambda=0.2$ of Fig. \ref{figvw} to $\lambda=0$ in order
to show the behavior in this limiting case. We see that the curves
are qualitatively similar for the three values of $\lambda$. Indeed,
$\mathcal{C}$ is positive for the ``unphysical'' branch of weak
detonations discussed in the previous section, and is negative for
the ``physical'' branch, as expected. (We plotted the negative part
of the curve in gray dashes to emphasize the fact that
$\mathcal{C}<0$ actually corresponds to marginal stability.) For a
better comparison, we show the curves of $\eta$ vs $v_w$ (dotted
lines).  As is observed, the change of sign of $\mathcal{C}$ occurs
exactly at the velocity $v_{\mathrm{crit}}$ corresponding to maximum
friction $\eta_{\max}$ [cf. Fig. \ref{figvw}] (we shall check this
analytically). Furthermore, $\mathcal{C}$ (and, thus, $\Omega$) seems
to diverge at the Jouguet velocity $v_J$. However, the present
treatment breaks down near the Jouguet point. Below, we analyze this
behavior in detail.

\subsection{Range of instability}

It is easy to see, from Eq. (\ref{B}), that the numerator
$\mathcal{N}$ will always change sign at the velocity $v_w =
v_{\mathrm{crit}}$, as observed in Fig. \ref{figstab}. Indeed, notice
that the condition for vanishing $\mathcal{N}$ is the same as Eq.
(\ref{etamax}) corresponding to the point separating the two branches
of solutions. Moreover, from the approximation (\ref{Bapp}), we see
that $\mathcal{N}$ is negative for $v_w=1$ and positive at the
Jouguet point, since the quantities $\Delta v$ and $v_-^2-c_{s-}^2$
vanish at these opposite ends of the weak detonation interval (in
blue in Fig. \ref{figvmavme}).

In the limit $v_w\to 1$, the approximation (\ref{Bapp}) is not valid,
since the factors $\gamma_-^{-1}$ in Eq. (\ref{B}) vanish. For
$\lambda=0$, the gamma factors cancel out. However, for $\lambda\neq
0$, both $\mathcal{N}$ and $\mathcal{D}$ vanish. Nevertheless, it is
not difficult to see that, in this limit, we have $\mathcal{N}\sim
\gamma_-^{-3},\mathcal{D}\sim \gamma_-^{-1}$. Hence, the parameter
$\mathcal{C}$ vanishes in the limit $v_w\to 1$, as seen in the
central and right panels of Fig. \ref{figstab}. This is the only
qualitative difference between the cases $\lambda=0$ and $\lambda\neq
0$, and it does not affect the stability analysis.

\begin{figure}[bth] \centering \epsfysize=5cm \leavevmode
\epsfbox{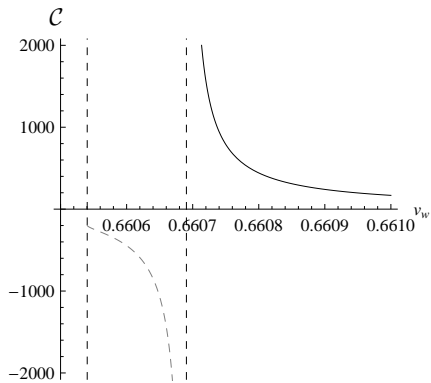} \caption{The same as Fig. \ref{figstab},
zooming in near $v_w=v_J$. Vertical dashed lines indicate the values $v_J$
and $v'_{\mathrm{crit}}$.}
\label{figstabzoom}
\end{figure}

Regarding the denominator $\mathcal{D}$, it is apparent in Eq.
(\ref{Aapp}) that it has a zero in the weak detonation range. This
zero is not at the Jouguet point. Indeed, for $v_-=-c_{s-}$, we have
$\mathcal{D}<0$. Hence, the divergence of $\Omega$ does not occur
exactly at the Jouguet point, but at a velocity
$v'_{\mathrm{crit}}>v_J$  (see Fig. \ref{figstabzoom}), although this
cannot be appreciated in Fig. \ref{figstab}. For
$v_w>v'_{\mathrm{crit}}$ we have $\mathcal{D}>0$, and the sign of the
stability parameter $\mathcal{C}$ is determined by the numerator
$\mathcal{N}$. As observed in Figs. \ref{figstab} and
\ref{figstabzoom}, the velocity $v'_{\mathrm{crit}}$ is very close to
the Jouguet velocity $v_J$. Below we shall see that this is the
general case.

Thus, the whole branch of weaker weak detonations
($v_{\mathrm{crit}}<v_w<1$) corresponds to $\mathrm{Re}(\Omega)<0$.
As already discussed, these solutions must be discarded, and we have
no solution with $\mathrm{Re}(\Omega)\neq 0$. Hence, for this
velocity range we have marginal stability and one should consider
non-linear perturbations. Nevertheless, the linear perturbation
result  suggests that these detonations will be stable. In contrast,
detonations in the velocity  range
$v'_{\mathrm{crit}}<v_w<v_{\mathrm{crit}}$ are unstable. This
constitutes almost the whole branch of stronger weak detonations,
since we have $v'_{\mathrm{crit}}\simeq v_J$.

It is important to notice that our treatment  breaks down for
velocities which are too close to the Jouguet point. In the first
place, for the unstable solutions on the right of the singularity,
the growth rate becomes higher and higher as $v_w$ approaches
$v'_{\mathrm{crit}}$. Hence, the small perturbation calculation will
break down as the characteristic growth time becomes much shorter
than the scale which characterizes the dynamics. This happens for
$\mathrm{Re}(\Omega)^{-1}\ll d$ (equivalently, for $\mathcal{C}\gg
1$). Therefore, the actual value of $\Omega$ may have a natural
cutoff. Nevertheless, even if $\mathrm{Re}(\Omega)$ does not diverge,
we expect a strong instability at $v_w= v'_{\mathrm{crit}}$.

For the solutions on the left of the singularity the situation is
much less clear. Here, we have $\mathrm{Re}(\Omega)<0$, which means
that the linear stability analysis is inconclusive. However, as we
have already discussed, the approach of considering perturbations
from a constant solution breaks down as we approach the Jouguet
point. Therefore, instead of considering non-linear perturbations,
one should consider perturbations around the appropriate fluid
profile. It is perfectly possible that such a calculation will give
linearly unstable modes. This would match smoothly the behavior for
$v_w> v'_{\mathrm{crit}}$.

We shall now estimate the value of $v'_{\mathrm{crit}}$ and analyze
how close it is to $v_J$ in the general case. To begin with, we can
see, from Eqs. (\ref{Aapp}) and (\ref{Bapp}), that
$v'_{\mathrm{crit}}$ is much closer to $v_J$ than
$v_{\mathrm{crit}}$. More explicitly, we can see that the
corresponding value of $v_-$ is much closer to $c_{s-}$. Indeed,
taking into account that  $\beta_-\gg 1$, we see that $\mathcal{N}$
vanishes for
 \beg \label{vcritapp0}
  \fr{|v_-|-c_{s-}}{c_{s-}}\approx \fr{1}{2}\fr{1}{s_-}\left(-\fr{\pa
  F_{\mathrm{dr}}}{\pa T_-}\right)\fr{\Delta v/v_+}{ F_{\mathrm{dr}}/w_-}.
 \en
while $\mathcal{D}$ vanishes for
 \beg \label{vcritpriapp0}
  \fr{|v_-|-c_{s-}}{c_{s-}}\approx \fr{1}{2}\fr{1}{s_-}\left(-\fr{\pa
  F_{\mathrm{dr}}}{\pa T_-}\right).
 \en
Thus, we see that, for the case of $v_{\mathrm{crit}}$, the value of
$v_-$ will lie somewhere between $c_{s-}$ and $1$, but cannot be very
close to $c_{s-}$, since at the Jouguet point we have a maximum
$\Delta v$ and  a minimum $F_{\mathrm{dr}}$. Moreover,
$F_{\mathrm{dr}}/w_-$ can be very small, since the reheating is
maximum at the Jouguet point\footnote{For instance, for the bag EOS
we have $F_{\mathrm{dr}}/w_-\sim\baL (1-T_-^2T_+^2/T_c^4)$, while,
estimating $\Delta v$ near the Jouguet point we have, from Eqs.
(\ref{fdrbag}-\ref{vj}), $\Delta v\sim \sqrt{\baL}$. Hence, we have
$F_{\mathrm{dr}}/w_-\ll \Delta v/v_+$.}.
On the other hand, the rhs of Eq. (\ref{vcritpriapp0}) is suppressed
by a factor $(F_{\mathrm{dr}}/w_-) /(\Delta v/v_+)$ with respect to
the rhs of Eq. (\ref{vcritapp0}). Hence, $v_-$ will be much closer to
$c_{s-}$ in this case. Consequently, $v'_{\mathrm{crit}}$ will be
much closer to $v_J$ than $v_{\mathrm{crit}}$.

\begin{figure}[bth] \centering \epsfysize=5cm \leavevmode
\epsfbox{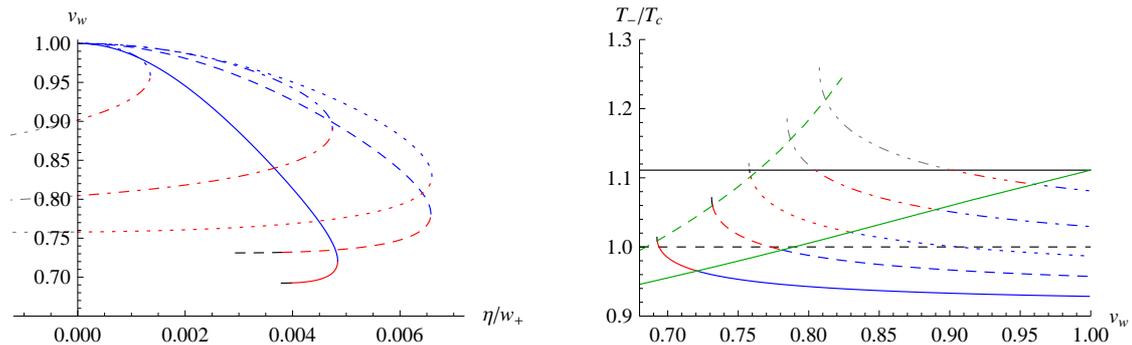} \caption{Left panel: The wall velocity for $\lambda=0$ as a function
of $\eta$, for $T_N/T_c=0.9$ and
$\baL=0.05$ (solid), $0.1$ (dashed), $0.15$ (dotted), $0.22$ (dot-dashed), and $0.3$ (dot-dot-dashed).
Right panel: The reheating as a function of the wall velocity, corresponding
to the curves of the left panel. Blue segments indicate  stable solutions and
red segments indicate  unstable solutions. Black segments indicate the region
where our stability analysis breaks down, and grey segments indicate
not realizable solutions which require negative values of $\eta$.
In the right panel,  the points corresponding to
$v'_{\mathrm{crit}}$ are indicated by a dashed green line, and those corresponding to $v_{\mathrm{crit}}$
by a solid green line. The horizontal dashed line indicates the value $T_-=T_c$, and the solid line indicates
the value $T_-=T_c^2/T_N$, for which the driving force vanishes.} \label{figtme}
\end{figure}

For a given model, we can estimate the value of $v_-$ corresponding
to $v_w=v'_{\mathrm{crit}}$ from Eq. (\ref{vcritpriapp0}). For the
bag EOS, this gives
 \beg \label{vcritpriapp}
  \fr{|v_-|-c_{s-}}{c_{s-}}\approx
  \frac12\fr{L}{w_-}\fr{T_-^2T_+^2}{T_c^4}
  \sim \fr{\baL}2.
 \en
Besides, since the Jouguet point is a minimum of $|v_+|$ vs $|v_-|$
(see Fig. \ref{figvmavme}),  we have $|v_+|-v_J\sim(|v_-|-c_{s-})^2$.
Hence, according to Eq. (\ref{vcritpriapp}), we have
 \beg \label{vcritprivj}
 v'_{\mathrm{crit}}-v_J\sim (\baL/2)^2.
 \en
The parameter $\baL$ is bounded by $1$, and in most cases we have
$\baL\ll 1$. Moreover, a large $\baL$ hinders the existence of
detonations, due to the hydrodynamical obstruction caused by the
release of latent heat. As can be observed in Fig. \ref{figtme} (left
panel), detonations cannot exist at all if the latent heat is too
large. How large, depends on the amount of supercooling (see Ref.
\cite{ms09}). For the case considered in the figure, i.e., for a
supercooling of $T_N/T_c=0.9$, the limit is $\baL\simeq 0.35$. This
hydrodynamic effect is most important near the Jouguet point, where
the reheating is maximum. For large enough latent heat, $T_-$
surpasses the critical temperature (see Fig. \ref{figtme}, right
panel). Moreover, the reheating may cause the driving force to become
negative. Of course, this means that such a solution is unreachable,
as it would require a negative friction (this can be observed in the
left panel of Fig. \ref{figtme}). For the approximation
(\ref{fdrbag}), this happens for $T_-=T_c^2/T_N$. As a consequence,
for large enough latent heat, the velocity $v'_{\mathrm{crit}}$,
being very close to the Jouguet point, will not be achieved. In the
case of Fig. \ref{figtme}, the existence of the Jouguet point
requires $\baL\lesssim 0.15$. Thus, for the cases where $v_J$ and
$v'_{\mathrm{crit}}$ exist we have, according to Eq.
(\ref{vcritprivj}), $v'_{\mathrm{crit}}-v_J\lesssim 0.005$.

\subsection{Bubble expansion, detonations, and runaway walls}

As already mentioned, hydrodynamic instabilities may determine, in
the case of multiple solutions, which one will be realized during a
phase transition. In principle, the fact that a solution is unstable
does not imply that it will not be realized. For instance, in the
case of an unstable deflagration, the wall is unstable under
{corrugations} on a characteristic scale $\lambda_c$, while long
wavelengths, as well as short wavelengths, are stable
\cite{link,kaj,stabdefla}. Thus, a bubble will grow to a size $\sim
\lambda_c$ before these corrugation instabilities can destabilize the
wall.

The case of detonations is certainly different. As we have seen, an
unstable detonation is unstable at all wavelengths with similar
growth rates. As a consequence, such a configuration, if ever formed
during bubble expansion, will decay immediately, either to some of
the stable configurations, such as a weaker weak detonation, or to a
runaway solution. Thus, weak detonations of the lower velocity branch
will most likely never be realized in a phase transition. In
particular, if the ultra-relativistic friction is small enough, there
will be no detonations at all, as in the cases represented in red in
Fig. \ref{figvw}.

As discussed in Sec. \ref{stationary},  detonations may coexist with
runaway solutions for certain ranges of parameters. For instance, as
we decrease the friction parameter $\eta$, the runaway solution
appears before the stationary solution ceases to exist (see Fig.
\ref{figvw}). This raises the question whether the detonation becomes
unstable when the runaway solution appears. According to the linear
perturbation theory, the answer seems to be no. As discussed above,
although inconclusive, the analysis suggests that the detonation
solution is stable in the range $v_{\mathrm{crit}}<v_w<1$. There is
no reason, thus, for the wall to run away, as long as a detonation
solution exists. Indeed, the runaway solution requires highly
ultra-relativistic velocities, i.e., an extremely high gamma factor
$\gamma_w$, while a detonation will most likely have $\gamma_w\sim
1$.

\subsection{Cosmological implications}

The unstable growth of a bubble wall may in principle have
interesting cosmological consequences. For instance, in the case of
deflagrations, instabilities under corrugations of the wall may lead
to dendritic growth \cite{fa90,link}. Such a complex dynamics may
have effects on electroweak baryogenesis \cite{kf92,a97}, magnetic
field generation \cite{soj97}, and gravity wave formation
\cite{stabdefla}. However, the case of unstable detonations is quite
different, since large wavelengths are as unstable as short
wavelengths. As we have discussed, such detonations will probably not
be formed at all, rather than forming and then decaying. Hence, in
this case there will not be such interesting effects. On the
contrary, any mechanism of relic generation which relies on stronger
weak detonation solutions will be negatively affected. Here we
discuss a couple of examples.

Being supersonic, detonation fronts are important for the generation
of gravitational waves. In this field, a Jouguet detonation has often
been assumed for the calculation of the gravitational wave background
from a cosmological phase transition. The Jouguet velocity depends
only on thermodynamical parameters and, for a given EOS, is easy to
calculate [see, e.g., Eq. (\ref{vj})]. However, if the friction force
is taken into account, the wall velocity will depend also on friction
parameters, and the solution will not necessarily be a Jouguet
detonation. Moreover, we have argued that this particular solution is
quite unlikely. In the first place, due to strong hydrodynamics, the
Jouguet velocity may just be unreachable for positive friction. Even
if the Jouguet point exists, it corresponds to the lower end of the
lower velocity branch of weak detonations, which behaves unphysically
as the parameters are varied. Unfortunately, our stability analysis
breaks down near the Jouguet point. However, as we have argued, the
fact that the growth rate of the instabilities becomes very large
near this point suggests that the Jouguet solution is probably
unstable as well\footnote{In any case, we have seen that previous
results claiming that the Jouguet detonation is stable \cite{r96} are
not valid.}. We remark, however, that weaker detonations, although
causing less disturbances in the fluid, have higher velocities and,
hence, may cause stronger gravitational waves (see, e.g.,
\cite{lm11}).

An interesting feature of detonations is the fact that the reheating
behind the wall may surpass the critical temperature (provided that
the latent heat is high enough for the given amount of supercooling).
An application of this fact is the interesting idea of supersonic
electroweak baryogenesis \cite{cn12}. In this scenario, small bubbles
of the symmetric phase nucleate in the superheated broken-symmetry
phase behind the wall. Baryogenesis occurs at the walls of these
small bubbles, which move slowly with respect to the fluid. Thus, the
necessary conditions for baryogenesis are fulfilled, and the required
baryon asymmetry is generated for reasonable values of the
parameters, although there are some restrictions. One of them is the
fact that the superheating must be strong enough for the symmetric
bubbles to fill a sizeable fraction of space.

In this respect, the analysis of Ref. \cite{cn12} lacks an important
feature of bubble growth, namely, the friction force. Indeed, the
wall velocity is left as a free parameter, and the temperature $T_-$
is calculated as a function of $v_w$. This is always possible, since
the reheating is given exclusively by hydrodynamics. As we have seen,
the fluid variables $T_-,v_-$ only depend on $T_N$ and $v_w$ (besides
the EOS parameters). In the general case, these relations can be
obtained from Eqs. (\ref{EM3.1}-\ref{EM3.2}). For the bag EOS, $T_-$
and $v_-$ are given by Eqs. (\ref{tme2}) and (\ref{steinhardt}),
respectively, as a function of $v_w$, $T_N/ T_c$, and $\baL$.
Moreover, a weak detonation profile like that of Fig. \ref{figdeto}
will always exist for each wall velocity in the range $v_J<v_w<1$.
However, if the velocity is not left as a free parameter but is
calculated, this interval will be reduced, due both to physical
impossibility and hydrodynamic instabilities. Unfortunately, the
reheating is higher for solutions which  are closer to the Jouguet
point. Therefore, most of the interesting solutions will belong to
the unstable branch.

Consider the plots of Fig. \ref{figtme}. The curves of $T_-$ vs. $v_w$
(right panel) can be plotted without knowing the friction force. We have
$T_->T_c$ for wide ranges of values of $\baL$ and $v_w$ (namely, all the
curves or parts of curves which lie above the horizontal dashed line).
However, if we actually calculate the velocity as a function of the
friction, we know that a part of each curve (the gray segments) corresponds
to velocities which cannot be achieved at all for any positive friction
parameter. Besides, another part of each curve corresponds to unstable
solutions (the red segments). Only the lower parts of the curves (the blue
segments) corresponds to stable detonations. Notice that, in fact, the upper
physical bound on $T_-$ (the horizontal solid line) is never reached for
stable solutions, which give values of $T_-$ below the solid green line.
This reduces the region of realizable solutions with $T_->T_c$ to the
triangular region between this line and the horizontal dashed line.

\begin{figure}[bth] \centering \epsfysize=8cm \leavevmode
\epsfbox{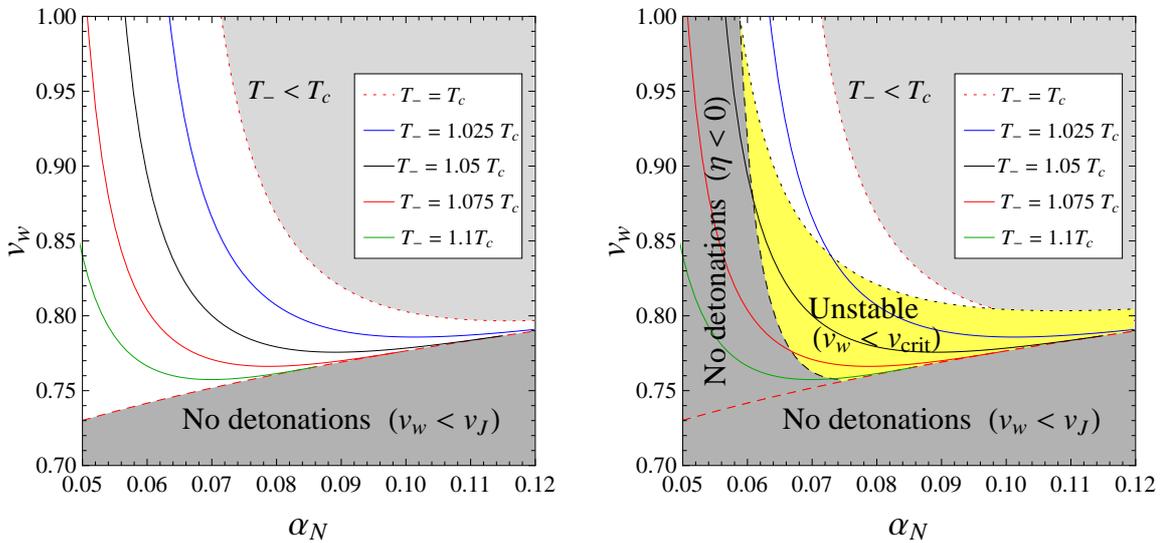} \caption{Region in the ($\alpha_N,v_w$)-plane for which $T_->T_c$, for $\baL=0.15$.
The red dashed line corresponds to $v_w=v_J$.
In the right panel, the black dashed line corresponds to solutions with $\eta=0$,
and the black dotted line corresponds to $v_w=v_{\mathrm{crit}}$.}
\label{figzonas}
\end{figure}

These restrictions on the reheating will certainly shrink the
available parameter space for the supersonic baryogenesis scenario,
since the symmetric bubbles need a considerable amount of
superheating to nucleate. It is out of the scope of the present work
to calculate the nucleation of symmetric bubbles, which would require
considering a specific model. In order to appreciate the restriction
on the scenario discussed in Ref. \cite{cn12}, we consider, as in
that work,  the detonation region in the ($\alpha_N,v_w$)-plane for
the bag EOS\footnote{In the notation of Ref. \cite{cn12}, we have
$a_+\equiv a$, $a_-\equiv a(1-\baL)$, and the value $\baL=0.15$
corresponds to $a_-/a_+=0.85$.}, for  $\baL=0.15$ (see Fig.
\ref{figzonas}). Here, the variable $\alpha_N=L/(4aT_N^4) $ is the
same as the variable $\alpha$ defined in Eq. (\ref{alfa}) (since for
detonations we have $T_+=T_N$). Thus, the left panel of Fig.
\ref{figzonas} coincides with Fig. 3 of Ref. \cite{cn12}. The region
in white corresponds to detonations for which the reheating exceeds
the critical temperature. We see that solutions with higher values of
$T_-$ are closer to the Jouguet point.

In the right panel of Fig. \ref{figzonas}, we have incorporated the
restrictions arising from the calculation of $v_w$ and from the stability
analysis. We considered the case $\lambda=0$, which is the most favorable
for detonations (for non-vanishing $\lambda$, runaway solutions appear and
detonations get reduced to smaller regions of parameter space). We see that
the region corresponding to $T_->T_c$ is significantly reduced, since most
of the detonations in this region are unphysical or unstable.  Notice that
the region corresponding to $T_-<T_c$ is also affected, but the reduction of
parameter space in this case is insignificant. This can be appreciated
already in the right panel of Fig. \ref{figtme}.

\section{Conclusions} \label{conclu}

The possible cosmological consequences of a first-order phase transition
depend on the dynamics associated to the motion of bubble walls. It is well
known that several hydrodynamic solutions are in principle possible for the
propagation of these phase-transition fronts. In particular, for the steady
state motion we may have weak or Jouguet detonations, as well as weak,
Jouguet or strong deflagrations. Nevertheless, some of these solutions are
in fact unstable. The stability analysis is thus important in order to
determine which of the propagation modes will be actually realized in a
phase transition.

In this work we have studied the stability of detonations under small
perturbations of the fluid and the interface. This paper is a sequel
of a similar study for the case of deflagrations \cite{stabdefla}.
The treatment of detonations is quite simpler, due to the fact that
the incoming fluid (in the reference frame of the wall) is
supersonic. Thus, the fluid in front of the wall is not affected by
the phase-transition front. This has two practical consequences for
the calculation. On the one hand, for the stationary case, the
incoming fluid velocity is given by the wall velocity, $v_+=-v_w$,
and the temperature in front of the wall is given by the boundary
condition $T_+=T_N$. On the other hand, perturbations originated at
the wall cannot grow in front of it. This simplifies the treatment of
the stability, since only perturbations of the outgoing fluid must be
considered (i.e., perturbations which vanish in front of the wall)
\cite{abney}. These simplifications allowed us to obtain analytical
and model-independent results.

Before the stability analysis, we have discussed in detail the
detonation solutions. It is well known that strong detonations are
hydrodynamically forbidden \cite{s82}. Weak detonations can exist in
the velocity range $v_J<v_w<1$, while the Jouguet detonation
corresponds to the case $v_w=v_J$. The Jouguet velocity $v_J$ depends
on the parameters of the equation of state, and is only determined by
the temperature $T_N$. However, the precise value of $v_w$ within
this interval will be determined by the balance of the driving and
friction forces, and will further depend on friction parameters. Due
to strong hydrodynamic effects, it turns out that some velocities in
this range may not be realized. Besides, the resulting velocity turns
out to be a double valued function of the parameters, and the bubble
wall will need to ``choose'' which of the two possible hydrodynamic
configurations to adopt.

Thus we have two branches of weak detonation solutions. One of them, with
velocities in the range $v_{\mathrm{crit}}<v_w<1$, where $v_{\mathrm{crit}}$
is given by Eq. (\ref{etamax}), corresponds to weaker weak detonations. The
other branch has velocities in the range $v_J<v_w<v_{\mathrm{crit}}$ and
corresponds to stronger weak detonations. This branch behaves rather
unphysically with the parameters. Although weak detonations are generally
believed to be stable \cite{hkllm}, numerical simulations \cite{ikkl94}
suggest that these stronger solutions are not.

We have applied the standard stability analysis for fluid interfaces
\cite{landau} to the case of relativistic detonation fronts in a
phase transition. We have pointed out that this analysis breaks down
at the Jouguet point. This is because the approach considers
perturbations around constant velocity and temperature, which is not
the case of the Jouguet detonation profile. As a consequence,
previous results on the stability of Jouguet detonations
\cite{abney,r96} are not valid. We have shown that  weak detonations
of the lower velocity branch are generally unstable under linear
perturbations at all wavelengths. More specifically, weak detonations
are unstable in the range $v_{\mathrm{crit}}'<v_w<v_{\mathrm{crit}}$,
where $v_{\mathrm{crit}}'$ is roughly given by Eq.
(\ref{vcritpriapp0}). Below $v_{\mathrm{crit}}'$, our approach breaks
down.  Nevertheless, as we have seen, $v_{\mathrm{crit}}'$ is very
close to the Jouguet velocity $v_J$. Therefore, it is quite unlikely
that the actual value of the wall velocity (taking into account the
friction) will fall in this very small interval. Regarding the higher
velocity branch, the linear perturbation analysis is not conclusive,
but we have argued that these solutions will probably be stable. This
means that, in the case of coexistence of a detonation and a runaway
solution, it is the stationary solution the one which will be
realized.

Our main result is, thus, that the branch of weak detonations which are
closer to the Jouguet point will not be realized during a phase transition.
Unfortunately, these solutions are cosmologically interesting due to their
strong disturbance of the fluid. We have discussed, in particular, how our
results affect a mechanism of electroweak baryogenesis with detonation walls
\cite{cn12}. This mechanism is based on the fact that the reheating behind
the wall may exceed the critical temperature, allowing the nucleation of
symmetric-phase bubbles in the superheated fluid inside the broken-symmetry
bubbles. The regions in parameter space for this scenario are significantly
reduced once unstable detonations are discarded.

\section*{Acknowledgements}

This work was supported by Universidad Nacional de Mar del Plata,
Argentina, grant EXA 607/12.


\begin{thebibliography}{99}


\bibitem{gw} For recent works, see, e.g.,
 C.~Caprini, R.~Durrer and G.~Servant,
Phys.\ Rev.\ D \textbf{77}, 124015 (2008) [arXiv:0711.2593 [astro-ph]];
 S.~J.~Huber and T.~Konstandin,
  JCAP {\bf 0809}, 022 (2008)
  [arXiv:0806.1828 [hep-ph]];
      S.~J.~Huber and T.~Konstandin,
  JCAP {\bf 0805}, 017 (2008)
  [arXiv:0709.2091 [hep-ph]].
 A.~Megevand,
  Phys.\ Rev.\  D {\bf 78} (2008) 084003
  [arXiv:0804.0391 [astro-ph]];
  T.~Kahniashvili, L.~Kisslinger and T.~Stevens,
  Phys.\ Rev.\  D {\bf 81}, 023004 (2010)
  [arXiv:0905.0643 [astro-ph.CO]].
      J.~Kehayias and S.~Profumo,
  JCAP {\bf 1003}, 003 (2010)
  [arXiv:0911.0687 [hep-ph]];
  J.~M.~No,
  Phys.\ Rev.\ D {\bf 84}, 124025 (2011)
  [arXiv:1103.2159 [hep-ph]];
     L.~Leitao, A.~Megevand and A.~D.~Sanchez,
  JCAP {\bf 1210}, 024 (2012)
  [arXiv:1205.3070 [astro-ph.CO]];
  M.~Hindmarsh, S.~J.~Huber, K.~Rummukainen and D.~J.~Weir,
  Phys.\ Rev.\ Lett.\  {\bf 112}, 041301 (2014)
  [arXiv:1304.2433 [hep-ph]].


\bibitem{gr01} For a review, see
  D.~Grasso and H.~R.~Rubinstein,
  Phys.\ Rept.\  {\bf 348}, 163 (2001)
  [arXiv:astro-ph/0009061].

\bibitem{vs94} A. Vilenkin and E.P.S.
    Shellard, {\it Cosmic Strings and Other Topological
    Defects} (Cambridge University Press, Cambridge, England,
    1994);
     A.~Vilenkin,
 Phys.\ Rept.\  {\bf 121}, 263 (1985).


\bibitem{ckn93} For reviews, see A.~G.~Cohen, D.~B.~Kaplan and
    A.~E.~Nelson,
Ann.\ Rev.\ Nucl.\ Part.\ Sci.\ \textbf{43}, 27 (1993)
[arXiv:hep-ph/9302210]; 
A.~Riotto and M.~Trodden, 
Ann.\ Rev.\ Nucl.\ Part.\ Sci.\ \textbf{49}, 35 (1999)
[arXiv:hep-ph/9901362]. 



\bibitem{w84}
E.~Witten, 
Phys.\ Rev.\ D \textbf{30}, 272 (1984);
G.~M.~Fuller, G.~J.~Mathews and C.~R.~Alcock,
Phys.\ Rev.\ D \textbf{37}, 1380 (1988);
J.~H.~Applegate and C.~J.~Hogan,
Phys.\ Rev.\ D \textbf{31}, 3037 (1985);
H.~Kurki-Suonio,
Phys.\ Rev.\ D \textbf{37}, 2104 (1988);
J.~Ignatius, K.~Kajantie, H.~Kurki-Suonio and M.~Laine,
Phys.\ Rev.\ D \textbf{50}, 3738 (1994) [arXiv:hep-ph/9405336].


\bibitem{h95} A.~F.~Heckler,
Phys.\ Rev.\ D \textbf{51} (1995) 405 [arXiv:astro-ph/9407064];

\bibitem{ma05} A.~Megevand and F.~Astorga,
  Phys.\ Rev.\ D {\bf 71}, 023502 (2005)
  [hep-ph/0409321].


\bibitem{hidro} See, e.g., M.~Gyulassy, K.~Kajantie, H.~Kurki-Suonio
    and L.~D.~McLerran,
Nucl.\ Phys.\ B \textbf{237}, 477 (1984); 
H.~Kurki-Suonio,
Nucl.\ Phys.\ B \textbf{255}, 231 (1985); 
K.~Kajantie and H.~Kurki-Suonio,
Phys.\ Rev.\ D \textbf{34}, 1719 (1986); 
K.~Enqvist, J.~Ignatius, K.~Kajantie and K.~Rummukainen,
Phys.\ Rev.\ D \textbf{45}, 3415 (1992). 


\bibitem{ms09} A.~Megevand and A.~D.~Sanchez,
Nucl.\ Phys.\ B \textbf{820}, 47 (2009)  [arXiv:0904.1753 [hep-ph]].

\bibitem{ekns10}
        J.~R.~Espinosa, T.~Konstandin, J.~M.~No and G.~Servant,
        JCAP {\bf 1006}, 028 (2010)
        [arXiv:1004.4187 [hep-ph]];

\bibitem{lm11}
  L.~Leitao and A.~Megevand,
  Nucl.\ Phys.\ B {\bf 844}, 450 (2011)
  [arXiv:1010.2134 [astro-ph.CO]].

\bibitem{kn11}  T.~Konstandin and J.~M.~No,
  JCAP {\bf 1102}, 008 (2011)
  [arXiv:1011.3735 [hep-ph]].

\bibitem{ms12}  A.~Megevand and A.~D.~Sanchez,
  Nucl.\ Phys.\ B {\bf 865}, 217 (2012)
  [arXiv:1206.2339 [astro-ph.CO]].



\bibitem{micro} See, e.g.,
 M.~Dine, R.~G.~Leigh, P.~Y.~Huet, A.~D.~Linde and
D.~A.~Linde, 
Phys.\ Rev.\ D \textbf{46}, 550 (1992) [arXiv:hep-ph/9203203];
 B.~H.~Liu, L.~D.~McLerran and N.~Turok,
Phys.\ Rev.\ D \textbf{46}, 2668 (1992).
N.~Turok, 
Phys.\ Rev.\ Lett.\ \textbf{68}, 1803 (1992);
S.~Y.~Khlebnikov,
Phys.\ Rev.\ D \textbf{46}, 3223 (1992); 
P.~Arnold, 
Phys.\ Rev.\ D \textbf{48}, 1539 (1993) [arXiv:hep-ph/9302258];
G.~D.~Moore and T.~Prokopec,
Phys.\ Rev.\ D \textbf{52}, 7182 (1995) [arXiv:hep-ph/9506475];
Phys.\ Rev.\ Lett.\ \textbf{75}, 777 (1995) [arXiv:hep-ph/9503296];
P.~John and M.~G.~Schmidt,
Nucl.\ Phys.\ B \textbf{598}, 291 (2001) [Erratum-ibid.\ B
\textbf{648}, 449
(2003)]; 
G.~D.~Moore,
JHEP \textbf{0003}, 006 (2000). 

\bibitem{bm09}   D.~Bodeker and G.~D.~Moore,
  JCAP {\bf 0905}, 009 (2009)
  [arXiv:0903.4099 [hep-ph]].

\bibitem{hs13}
  S.~J.~Huber and M.~Sopena,
  arXiv:1302.1044 [hep-ph].

\bibitem{ariel13}   A.~Megevand,
  JCAP {\bf 1307}, 045 (2013)
  [arXiv:1303.4233 [astro-ph.CO]].



\bibitem{s82} P.~J.~Steinhardt,
Phys.\ Rev.\ D \textbf{25}, 2074 (1982). 

\bibitem{l94} M.~Laine, 
Phys.\ Rev.\ D \textbf{49}, 3847 (1994) [arXiv:hep-ph/9309242].



\bibitem{hkllm}  P.~Y.~Huet, K.~Kajantie, R.~G.~Leigh, B.~H.~Liu and
    L.~D.~McLerran,
Phys.\ Rev.\ D \textbf{48}, 2477 (1993) [arXiv:hep-ph/9212224].

\bibitem{stabdefla}   A.~Megevand and F.~A.~Membiela,
  arXiv:1311.2453 [astro-ph.CO].


\bibitem{landau} L. D. Landau and E. M. Lifshitz, \textit{Fluid
    Mechanics}
(Pergamon Press, New York, 1989).

\bibitem{link}   B.~Link,
  Phys.\ Rev.\ Lett.\  {\bf 68}, 2425 (1992).



\bibitem{ikkl94} J.~Ignatius, K.~Kajantie, H.~Kurki-Suonio and
    M.~Laine,
Phys.\ Rev.\ D \textbf{49}, 3854 (1994); 
  H.~Kurki-Suonio and M.~Laine,
  Phys.\ Rev.\  D {\bf 51}, 5431 (1995)
  [arXiv:hep-ph/9501216];
    H.~Kurki-Suonio and M.~Laine,
  Phys.\ Rev.\ D {\bf 54}, 7163 (1996)
  [hep-ph/9512202].



\bibitem{abney}   M.~Abney,
  Phys.\ Rev.\ D {\bf 49}, 1777 (1994)
  [astro-ph/9305021].

\bibitem{r96}   L.~Rezzolla,
  Phys.\ Rev.\ D {\bf 54}, 1345 (1996)
  [astro-ph/9605033].



\bibitem{quiros}    M.~Quiros,
  arXiv:hep-ph/9901312.


\bibitem{gw81} A.~H.~Guth and E.~J.~Weinberg,
Phys.\ Rev.\ D \textbf{23}, 876 (1981). 

\bibitem{linde}
A.~D.~Linde, 
Nucl.\ Phys.\ B \textbf{216}, 421 (1983) [Erratum-ibid.\ B
\textbf{223}, 544
(1983)]; 
Phys.\ Lett.\ B \textbf{100}, 37 (1981). 

\bibitem{ah92} G.~W.~Anderson and L.~J.~Hall,
Phys.\ Rev.\ D \textbf{45}, 2685 (1992). 

\bibitem{mege}   A.~M\'egevand,
  Int.\ J.\ Mod.\ Phys.\ D {\bf 9}, 733 (2000)
  [hep-ph/0006177];
  A.~Megevand,
  Phys.\ Rev.\ D {\bf 64}, 027303 (2001)
  [hep-ph/0011019];
  A.~Megevand,
  Phys.\ Rev.\ D {\bf 69}, 103521 (2004)
  [hep-ph/0312305];
  A.~Megevand and A.~D.~Sanchez,
  Phys.\ Rev.\ D {\bf 77}, 063519 (2008)
  [arXiv:0712.1031 [hep-ph]];
  A.~Megevand and A.~D.~Sanchez,
  Nucl.\ Phys.\ B {\bf 825}, 151 (2010)
  [arXiv:0908.3663 [hep-ph]].



\bibitem{kaj}   K.~Kajantie,
  Phys.\ Lett.\ B {\bf 285}, 331 (1992).

\bibitem{fa90}   K.~Freese and F.~C.~Adams,
  Phys.\ Rev.\ D {\bf 41}, 2449 (1990).

\bibitem{kf92}
  M.~Kamionkowski and K.~Freese,
  Phys.\ Rev.\ Lett.\  {\bf 69}, 2743 (1992)
  [hep-ph/9208202].

\bibitem{a97}
  M.~Abney,
  Phys.\ Rev.\ D {\bf 55}, 582 (1997)
  [hep-ph/9606476].

\bibitem{soj97}    G.~Sigl, A.~V.~Olinto and K.~Jedamzik,
  Phys.\ Rev.\ D {\bf 55}, 4582 (1997)
  [astro-ph/9610201].


\bibitem{cn12}   C.~Caprini and J.~M.~No,
  JCAP {\bf 1201}, 031 (2012)
  [arXiv:1111.1726 [hep-ph]].


\end{thebibliography}
\end{document}